%% file: chi26-107.tex
\documentclass[sigconf]{acmart}

\usepackage[utf8]{inputenc}
\usepackage{graphicx}

\usepackage{subcaption , acmart-taps}
\usepackage{enumitem}
\usepackage{array,booktabs}

\newcolumntype{L}[1]{>{\raggedright\arraybackslash}p{#1}}
\newcolumntype{Y}{>{\raggedright\arraybackslash}X}

\AtBeginDocument{}
\newcommand{\q}[1]{\emph{``#1''}}

\copyrightyear{2026}
\acmYear{2026}
\setcopyright{cc}
\setcctype{by}
\acmConference[CHI '26]{Proceedings of the 2026 CHI Conference on Human Factors in Computing Systems}{April 13--17, 2026}{Barcelona, Spain}
\acmBooktitle{Proceedings of the 2026 CHI Conference on Human Factors in Computing Systems (CHI '26), April 13--17, 2026, Barcelona, Spain}
\acmPrice{}
\acmDOI{10.1145/3772318.3790370}
\acmISBN{979-8-4007-2278-3/2026/04}

\begin{document}

\title{Actor’s Note: Examining the Role of AI-Generated Questions in Character Journaling for Actor Training}

\author{Sora Kang}
\email{sorakang@snu.ac.kr}
\affiliation{%
  \institution{Seoul National University}
  \city{Seoul}
  \country{Republic of Korea}
}

\author{Jaemin Zoh}
\email{jmzoh218@snu.ac.kr}
\affiliation{%
  \institution{Seoul National University}
  \city{Seoul}
  \country{Republic of Korea}
}

\author{Hyoju Kim}
\email{gywn0429@snu.ac.kr}
\affiliation{%
  \institution{Seoul National University}
  \city{Seoul}
  \country{Republic of Korea}
}

\author{Hyeonseo Park}
\email{tthnso@snu.ac.kr}
\affiliation{%
  \institution{Seoul National University}
  \city{Seoul}
  \country{Republic of Korea}
}

\author{Hajin Lim}
\email{hajin@snu.ac.kr}
\affiliation{%
  \institution{Seoul National University}
  \city{Seoul}
  \country{Republic of Korea}
}

\author{Joonhwan Lee}
\email{joonhwan@snu.ac.kr}
\affiliation{%
  \institution{Seoul National University}
  \city{Seoul}
  \country{Republic of Korea}
}



\begin{teaserfigure}
  \centering
  \includegraphics[width=\textwidth]{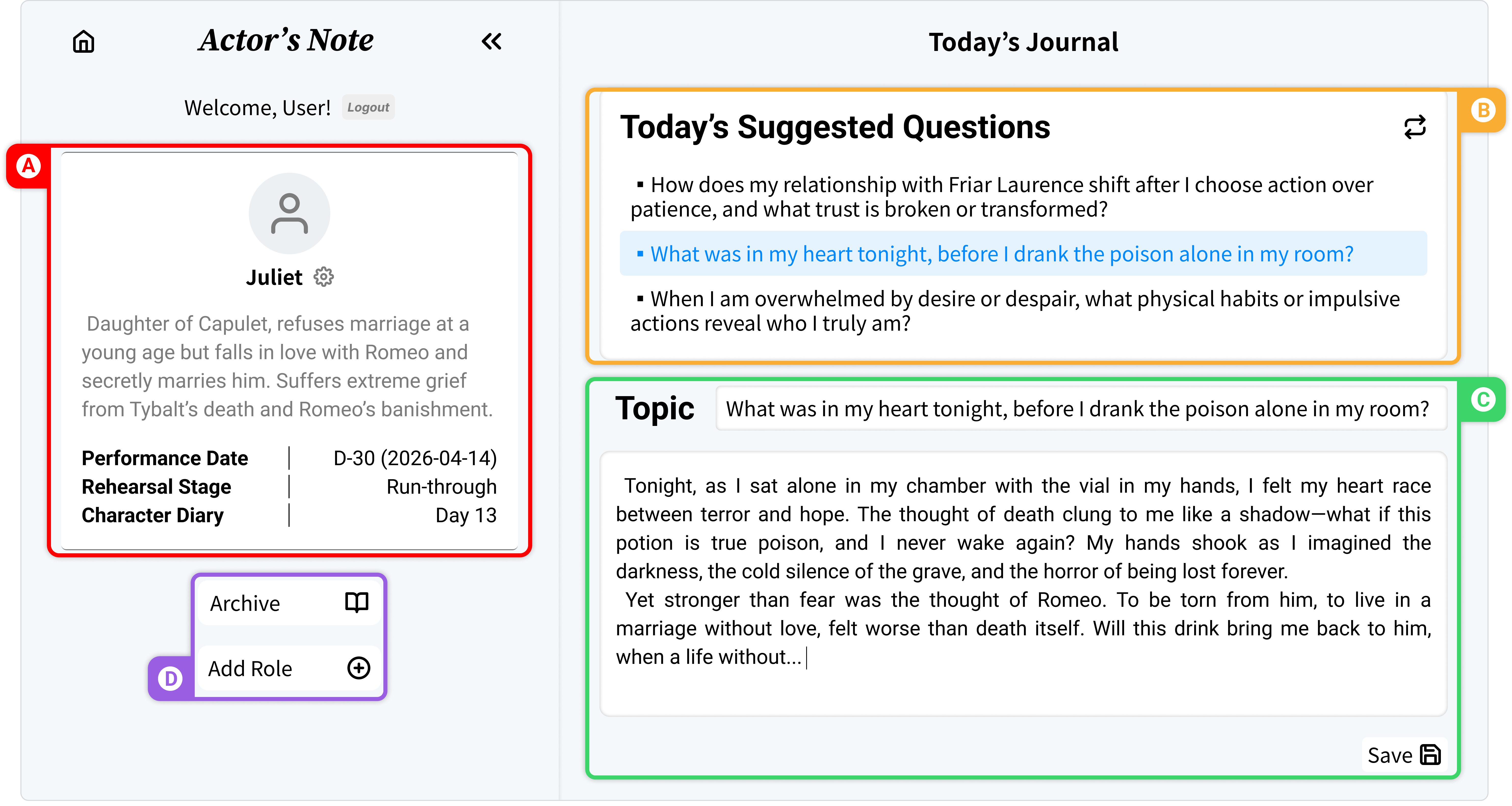}
  \caption{Actor's Note main interface showing (A) a character profile derived from script, role, rehearsal stage, and performance date; (B) three stage-aware suggested questions; (C) the journal text editor and (D) archive access and role addition.}
  \Description{Dashboard interface divided into four labeled regions. A character-profile panel appears in the upper-left as a block of text fields. To its right, three stacked rectangular boxes each contain one question. A large text-entry region spans the lower center of the screen. Along the right side, a tall scrollable archive panel lists prior items in rows. The four regions are visually separated by panel borders.}
  \label{fig:teaser}
\end{teaserfigure}

\renewcommand{\shortauthors}{Sora Kang et al.}

\begin{abstract}
\input{section/0Abstract}
\end{abstract}

\maketitle

\section{Introduction}
\input{section/1Introduction}

\section{Related Work}
\input{section/2RelatedWork}

\section{Formative Study}
\input{section/3FormativeStudy} 

\section{Actor's Note}
\input{section/4System}

\section{Method}
\input{section/5Method}

\section{Results}
\input{section/6Results}

\section{Discussion}
\input{section/7Discussion}

\section{Conclusion}
\input{section/8Conclusion}

\begin{acks}

We would like to thank the anonymous reviewers for their thoughtful and constructive feedback, which greatly helped improve the quality of this work. We are also immensely grateful to the anonymous actors who participated in our study; their artistic voices and dedicated engagement made this research possible.
This work was partly supported by Institute of Information \& Communications Technology Planning \& Evaluation (IITP) grant funded by the Korea government (MSIT) [NO.RS-2021-II211343, Artificial Intelligence Graduate School Program (Seoul National University)], SNU-Global Excellence Research Center establishment project, and Basic Science Research Program through the National Research Foundation of Korea (NRF) funded by the Ministry of Education (No. RS-2025-25421701).

\end{acks}

\bibliographystyle{ACM-Reference-Format}
\bibliography{chi26-107}

\input{section/appendix}

\end{document}

%% file: section/0Abstract.tex
Character journaling is a well-established exercise in actor training, but many actors struggle to sustain it due to cognitive burden, the blank page problem, and unclear short-term rewards. We reframe large language models not as co-authors but as maieutic partners—tools that guide reflection through context-aware questioning rather than producing text on behalf of the user. Based on this perspective, we designed Actor’s Note, a journaling tool that tailors questions to the script, role, and rehearsal phase while preserving actor agency. We evaluated the system in a 14-day crossover study with 29 actors using surveys, logs, and interviews. Results indicate that the tool reduced entry barriers, supported sustained reflection, and enriched character exploration, with participants describing different benefits when AI was introduced at earlier versus later rehearsal stages. This work contributes empirical insights and design principles for creativity-support tools that sustain reflective practices while preserving artistic immersion in performance training.

%% file: section/1Introduction.tex
Character journaling has long been a cornerstone of actor training~\cite{bruder2012practical, cremin2006connecting, stanley2022actor, writinginrole, locanto2025essential}. In pedagogical traditions grounded in Stanislavski's system and Method acting, writing and thinking in-role function as reflective methods through which actors adopt a character's perspective, articulate motivations, and imagine a life beyond the script~\cite{blatner1996acting, carnicke1999stanislavsky, chekhov2013actor, krasner1999strasberg, moore1984stanislavski}. Through journaling, actors expand backstories and explore emotions that may never explicitly appear on stage, yet critically inform their performance~\cite{hagen1991challenge, spaces2023actor, stanislavski2009actor, writinginrole}. It is therefore not mere documentation but a method for constructing an internally coherent character that can be reliably accessed under performance conditions~\cite{utahagen9questions, mitchell2008director, preeshl2016past}. This practice enables actors to inhabit a character's inner life with depth and authenticity.

Yet despite its recognized pedagogical value, sustaining character journaling in practice has been reported as difficult. First, the rehearsal process already imposes substantial cognitive and temporal demands, leaving little energy for additional reflective writing~\cite{blix2015professional, burgoyne1999impact, norrthon2023knowledge, taylor2016actor}. Second, many actors struggle with the \q{blank page} problem, a sense of not knowing what to write, which raises the barrier to starting and results in abandoned attempts~\cite{joyce2009blank, schultz1985writer}. Third, the effects of journaling on performance can feel abstract, as it is rarely clear how exactly a written practice translates into improved acting in the short term~\cite{sorensen2024always}, especially given the inherently interpretive and difficult-to-measure nature of acting performance itself. Finally, the practice is typically undertaken in isolation, without structured guidance or feedback from others, which further diminishes motivation~\cite{bascomb2019performing}. Together, these factors mean that even when actors acknowledge the benefits of journaling, they often fail to sustain it over time.

Recent advances in large language models (LLMs) provide new opportunities to address these challenges. LLMs have demonstrated strong capabilities in creativity-support tasks, including storytelling, script rewriting, and character analysis~\cite{begus2024experimental, branch2021collaborative, dayo2023scriptwriting, magee2024drama, dharaniya2023design}. However, applications remain largely creator-centric, focusing on authorship and content generation~\cite{chakrabarty2024creativity, coenen2021wordcraft}. Their potential within actor training—a practice centered on interpretation and character immersion—remains under-examined. The question is not whether AI can write for actors, but whether it can act as a training partner that lowers barriers while preserving autonomy and creativity.

To explore this gap, we designed and developed \textit{Actor’s Note}, an interactive, stage-aware journaling system powered by an LLM. \textit{Actor’s Note} scaffolds the journaling process by providing contextual topics and questions tailored to the actor’s role, script, and rehearsal stage. We evaluated the system in a 14-day, in-the-wild randomized crossover study with 29 actors to address the following research question: \textit{How can an AI-powered tool support actors’ character journaling practice, and what are its effects on their character understanding, acting confidence, creativity, and sense of agency?}

This paper makes the following contributions:
\begin{itemize}[leftmargin=*]
    \item \textbf{A Novel System for Creative Practice:} We present \textit{Actor’s Note}, an AI-assisted journaling tool that adapts a long-standing pedagogical practice into a computationally supported form, featuring a stage-aware, script-grounded prompting mechanism.
    \item \textbf{An Empirical Study of AI in Actor Training:} We provide findings from a 14-day, in-the-wild randomized crossover study with 29 actors, examining how journaling behaviors, perceived cognitive load, and reflective engagement differ between practice-as-usual freewriting and an AI-assisted condition.
    \item \textbf{Design Implications for Creativity Support Tools:} We derive principles for designing AI systems that scaffold, rather than automate, high-immersion creative and interpretive practices, framing AI as a maieutic partner while balancing structured support with user agency.
    \item \textbf{Pedagogical implications on timing of introduction:} From the crossover, we identify differential benefits by timing and translate these patterns into actionable guidance for integrating AI into actor training.
\end{itemize}

%% file: section/2RelatedWork.tex
This section reviews prior work on character journaling in actor training, LLMs in creative contexts, and reflection systems.

\subsection{Actor Training and Character Journaling}
Character journaling, often described as a form of \q{writing in role,} is a training method in which actors write from the first-person perspective of their characters~\cite{hancock1993character, noice1994example}. While Stanislavski did not prescribe journaling as a training method, his concepts of experiencing and the Magic If—which require actors to inhabit a character's inner life and imagine themselves in fictional circumstances~\cite{benedetti2004stanislavski, moore1984stanislavski, polanco2016stanislavski, stanislavski2013building, stanislavskij1986actor}—have inspired later pedagogical practices that incorporate writing in role as a means of deepening immersion and reflection~\cite{cremin2006connecting, ranzau2017drama, writinginrole}.

In this technique, actors produce diary- or letter-like texts from the standpoint of a character. It serves as a powerful tool for an actor's deep immersion into a character's inner world, facilitating the exploration of motivations, emotional histories, and nuanced perspectives that allow them to build authentic, lived-in portrayals~\cite{characterjournals2025, clare2016stanislavsky, lee2020writer}. However, despite its potential, the practice often fails to be fully utilized due to the absence of systematic guidance and the difficulty of sustaining reflective momentum. Stanislavski's system offered the philosophical foundation of inner truth and imaginative inquiry~\cite{stanislavski2013building, stanislavskij1986actor}, and later scholars have likewise tended to treat rehearsal journals as personal, implicit practices rather than codified protocols~\cite{kahan1985introduction, tamiolaki2024creating, writinginrole}. Consequently, while many contemporary actors keep rehearsal journals, methods and outputs remain largely unsystematized~\cite{merlin2013using}. In practice, actors face barriers such as cognitive load, time pressure, the \q{blank page} problem, and a lack of tangible feedback, which hinder sustained engagement~\cite{joyce2009blank, sorensen2024always}.

Furthermore, while empirical studies show that actor training improves cognitive perspective-taking (based on Theory of Mind) and communicative authenticity, these studies assess the aggregate effect of \q{training} and not isolate or validate the independent effect of journaling itself~\cite{edinborough2011developing, goldstein2011correlations, goldstein2012enhancing, mcdonald2020could}. Thus, character journaling remains a practice that is theoretically vital yet empirically and procedurally weak. This poses a pedagogical challenge: How can we systematically and sustainably support an actor's reflective process without losing the essential nature of personal inquiry? We frame this as an HCI problem of designing and evaluating \textit{what }to write, \textit{when}, and \textit{how}, and propose the LLM as a means to address it.

\subsection{LLMs as Creative and Performative Partners}
LLMs have increasingly been integrated into creative workflows, with growing evidence that they can interpret context and sustain coherent dialogue and logical reasoning~\cite{huang2022towards, liu2024speak, liu2025logical, yi2024survey, zhu2024large}. Prior work shows that these capabilities facilitate ideation and writing flow through mixed-initiative exchanges in which humans and models propose, revise, and critique~\cite{coenen2021wordcraft}; maintain long-range structure and coherence via hierarchical generation~\cite{mirowski2022cowriting, zhong2023memorybank}; support session-to-session continuity through long-term memory and reflection loops~\cite{liang2024self, park2023generative, shinn2023reflexion}; and enable exploration of alternatives through counterfactual reasoning. These forms of context-sensitive dialogue and reasoning have been validated most clearly in creative writing: given a narrative setup, models are able to propose alternative plot developments~\cite{tian2024large, wang2024storyverse, yuan2022wordcraft}, generate dialogue in varied styles~\cite{carrera2025nabokov, han2025stagewizard}, and summarize or restructure complex narratives~\cite{gero2022sparks, tian2024large, xie2023next}, thereby materially supporting the writing process~\cite{ahn2024transformative, chakrabarty2024creativity}. Collectively, these abilities provide a technical basis for addressing character journaling barriers such as the blank-page problem, single-interpretation lock-in, and lack of continuity.

Beyond writing support, these capabilities have also been explored on stage. Studies report systems that use LLMs for live audience interaction in comedy or as the \q{brain} of a digital performer for improvisation~\cite{branch2024designing, drago2025improvmate, mathewson2017improvised, rond2019improv}. Theatrical Language Processing positions the model as an improvisational partner to co-create scenes and seed new scripts in real time~\cite{kang2025theatrical}, and Mirowski et al. investigate enhancing audience experience in live comedy with LLMs~\cite{mirowski2025theater}. Yet this literature largely treats the model as a tool for performative output—scene generation and audience interaction—i.e., actors' external practice. Little research addresses LLMs as educational supports for actors' inner inquiry and training. We instead reposition the model as a maieutic partner and design question-centered scaffolding that helps actors, in the first person, structure motives, affective arcs, and perspectives.

\subsection{Digital Journaling and Reflection Systems}
The HCI community has accumulated extensive research on digital tools for journaling and self-reflection~\cite{bentvelzen2022revisiting, engin2011research, rieman1993diary}, with broad agreement that light scaffolding enhances the depth and persistence of reflection~\cite{engin2011research, jang2025journey, king2006ejournaling, li2024diaryhelper, wang2025designing}. For example, prompt-based scaffolding—templates, progressive prompts, example sentences, and domain-specific vocabulary cues—lowers the blank-page barrier and broadens the scope of thought~\cite{fleck2010reflecting, zhang2013prompts}, and review loops that summarize and highlight prior entries while surfacing next goals provide cross-session cues that accumulate reflection~\cite{epstein2015lived, li2010stage}. These design principles have increased users' self-awareness and supported consistent participation across domains such as well-being, education, and productivity~\cite{sawhney2018audio, zhang2016examining}. Building on these patterns, AI-augmented journaling has begun to appear: systems use conversational guidance to reframe or elaborate prompts, apply LLM-based summarization to surface follow-up questions, and offer context-aware reminders that adapt to recent activity and prior entries. Reports generally indicate higher self-awareness and better adherence relative to unguided journaling~\cite{nepal2024contextual, wang2025designing, zhou2025journalaide}. Whereas many systems emphasize text generation or elaboration, our system operationalizes a maieutic stance that elicits reflection rather than generating content.

Character journaling appears particularly well-suited to benefit from these strategies and technical aids, yet within our review scope, to our knowledge, there has been limited work that systematically integrates and evaluates them in the actor-training context. We therefore reinterpret these validated design assets for actor training: minimal intervention, stage-aware pacing, and contextual continuity utilizing AI. We retain scaffolding that deepens reflection while avoiding over-structuring that disrupts creative flow, adopting a question-driven maieutic stance that supports actors' self-reflection while preserving their authorship.

%% file: section/3FormativeStudy.tex
To ground our system design in actor practice, we conducted a one-week pilot of a web prototype. Our goals were to examine how AI-generated questions would be perceived, assess their utility for character journaling, and derive final system design requirements.

\begin{table*}[t]
  \renewcommand{\arraystretch}{0.5} 
  \caption{Formative study design traceability. Observed challenges from a one-week pilot are mapped to design implications and the concrete changes implemented in Actor's Note.}
  \label{tab:traceability}
  \centering
  \begin{tabular}{p{0.31\textwidth} p{0.31\textwidth} p{0.31\textwidth}}
    \toprule
    \textbf{Observed Challenge} & \textbf{Design Implication} & \textbf{Change in Actor's Note} \\
    \midrule
    Fundamental questions arriving too close to performance felt destabilizing. & Stage-aware scaffolding tuned to rehearsal phase; add guardrails near performance. & Stage and D-Day collected at onboarding; prompt context includes them. \\
    \addlinespace 
    Too many prompts at once created hesitation and cognitive load. & Pacing \& focus control with a small set of concise options. & Three question cards per session. \\
    \addlinespace
    Occasional need to bypass the suggested prompt to capture top-of-mind material. & Preserve agency; let users steer and ground questions in role/script. & Enable skip, edit, and select for questions to preserve agency and individuality. \\
    \addlinespace
    Desire for more variety across angles to sustain momentum. & Lightweight thematic scaffold to broaden exploration without over-structuring. & A five-category framework to provide continuity without over-structuring. \\
    \addlinespace
    Actors wanted to \q{just write there.} & Integrated notebook interaction. & In-place editor and an entry archive. \\
    \bottomrule
  \end{tabular}
\end{table*}

\subsection{Prototype}
We built a web-based prototype that generates questions from an uploaded script using the GPT-4o model with instruction-style prompting (system prompts and exemplars). Exemplars grounded in Uta Hagen's \textit{Nine Questions}~\cite{utahagen9questions} steered the model to adapt that framing to each role and the script's plot. The tool generated questions only and was accessible on mobile and desktop.

\begin{figure}[h]
  \centering
  \includegraphics[width=\linewidth]{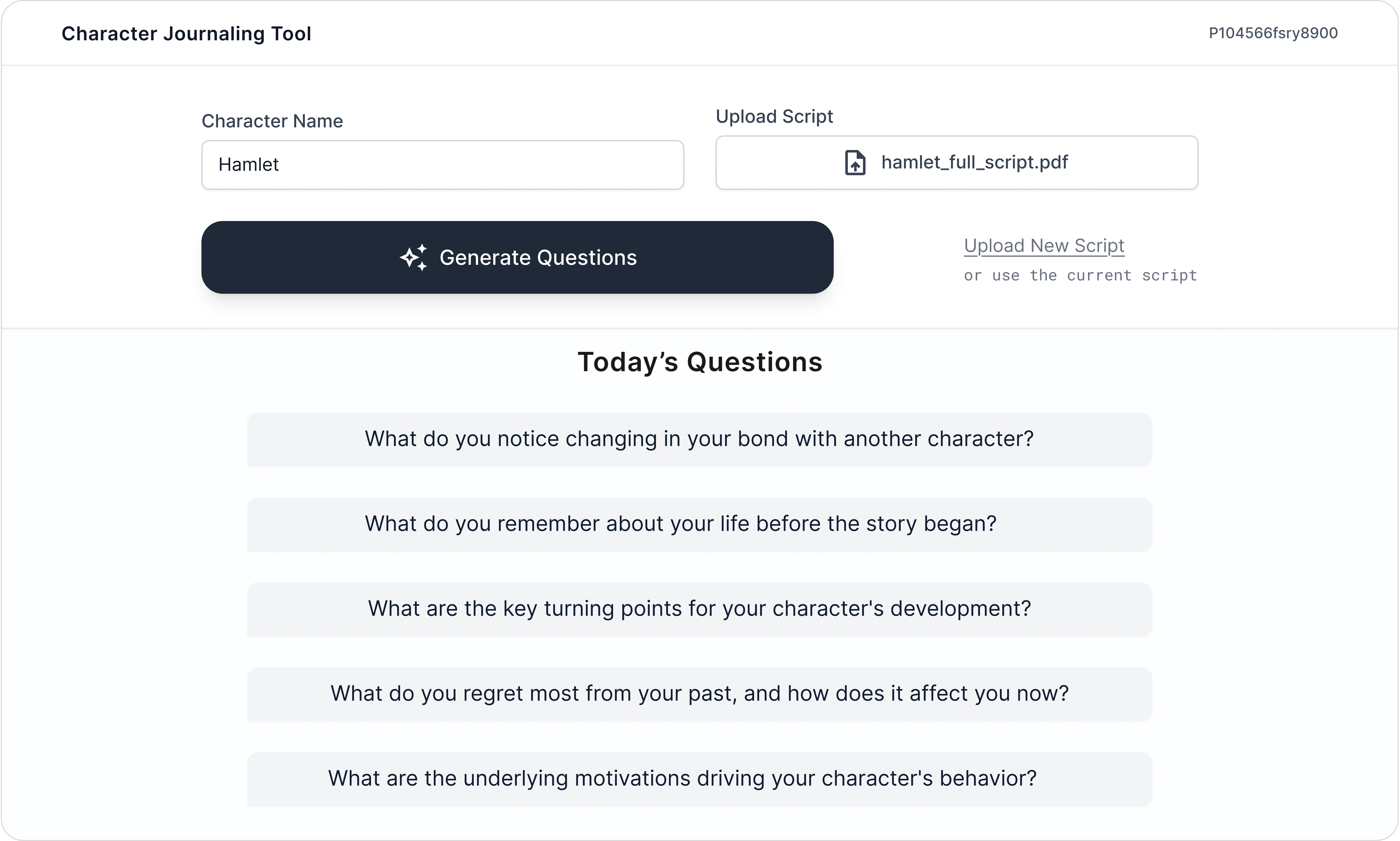}
  \caption{Early web prototype used in the formative study.}
  \Description{Early prototype page with a vertical layout. A large text box for inserting script text appears at the top. Beneath it are smaller input fields and a wide button. Below the button, a column of rectangular boxes displays generated questions, each occupying one row. No journaling editor or archive elements appear in this version.}
  \label{fig:prototype}
\end{figure}

\subsection{Procedure}
We deployed the prototype for one week with eight current or former actors ($N = 8$; 2--20+ years of experience). Recruitment used snowball sampling (min. 2 years experience). Participants received a link and instructions and were asked to write at least five entries during the week, guided by the suggested questions using their preferred tools (e.g., note-taking apps) or by hand; the prototype did not collect journal text. At week's end, we ran a two-hour semi-structured focus group to elicit experiences and surface design requirements. The session was audio-recorded, automatically transcribed, and then manually corrected. Two researchers conducted an inductive thematic analysis: we independently open-coded an initial subset, reconciled discrepancies through discussion to form a shared codebook, and then applied it to the remaining transcript. Given the small sample and single-session format, we prioritized consensus over formal inter-rater statistics. All participants provided informed consent; participation was voluntary and uncompensated, and all quotes anonymized.

\subsection{Findings}
All eight participants completed the one-week pilot. Based on end-of-week self-reports, five participants wrote every day and three wrote six out of seven days.

Participants consistently described three benefits of the AI-generated questions. First, they lowered the barrier to starting: several actors said the daily topics helped them start writing---\q{Having a topic each day made it easy to start writing,} noted P02. Second, questions also encouraged broader interpretation; P05 said, \q{I hadn't realized how central the uncle was to this character until the question made me look for it.} Third, the prompts unsettled fixed readings; P07 reflected, \q{I'd played this role before; the AI nudged me to reframe and then lock a different choice.}

At the same time, the pilot surfaced clear risks and limits. Timing mattered: roughly half the group cautioned that fundamental ``why'' questions, if introduced late in rehearsal or near performance, could destabilize settled choices---\q{Some questions were so basic that it took me a long time to answer; during a run, that would be confusing,} said P06. Several participants also experienced prompt overload when too many suggestions arrived at once, even as they wanted greater topical variety---\q{When a bunch of questions arrived together, I didn't know which to pick, but I still wanted more diverse angles,} said P03. Workflow friction was another theme: copying prompts into external tools interrupted flow, creating demand for in-place journaling. Finally, adherence to the session protocol---writing on the suggested topic---left some participants dissatisfied: \q{I had a topic I really wanted to write about that day, and I was disappointed when it didn't show up in the suggestions,} said P07.

Based on these observations, we distilled design principles for the next iteration: (i) stage-aware questioning aligned with rehearsal phases; (ii) pacing controls (max. 3 prompts); (iii) a five-category scaffold; and (iv) actor agency via skip/edit/select (see Table~\ref{tab:traceability}).

%% file: section/4System.tex
Based on the findings from our formative study, we developed \textit{Actor's Note}. We conceptualize \textit{Actor's Note} as a technology probe~\cite{hutchinson2003technology}, designed not as a definitive solution, but an instrument to explore how AI can augment the quality and sustainability of actors' character-journaling practice. Its design is guided by five goals drawn from those findings:

\begin{itemize}
    \item \textbf{(DG1) Minimize Cognitive Load:} Lower the barrier to initiating and maintaining the writing practice.
    \item \textbf{(DG2) Preserve Actor Agency:} Ensure actors remain in control of their creative and interpretive process.
    \item \textbf{(DG3) Ensure Stage-Appropriateness:} Adapt the nature of questions to the actor's rehearsal phase.
    \item \textbf{(DG4) Maintain Character Uniqueness:} Inspire exploration without prescribing a specific characterization.
    \item \textbf{(DG5) Enable Measurable Interaction:} Collect interaction logs to enable a quantitative analysis.
\end{itemize}

To embody these principles---especially minimal intervention---the core interaction is intentionally simple. Rather than writing for the actor, \textit{Actor's Note} functions as a partner: after initial script analysis, it surfaces three tailored questions at once to support a proactive and sustainable routine.

\begin{figure*}[t]
  \centering
  \includegraphics[width=\textwidth]{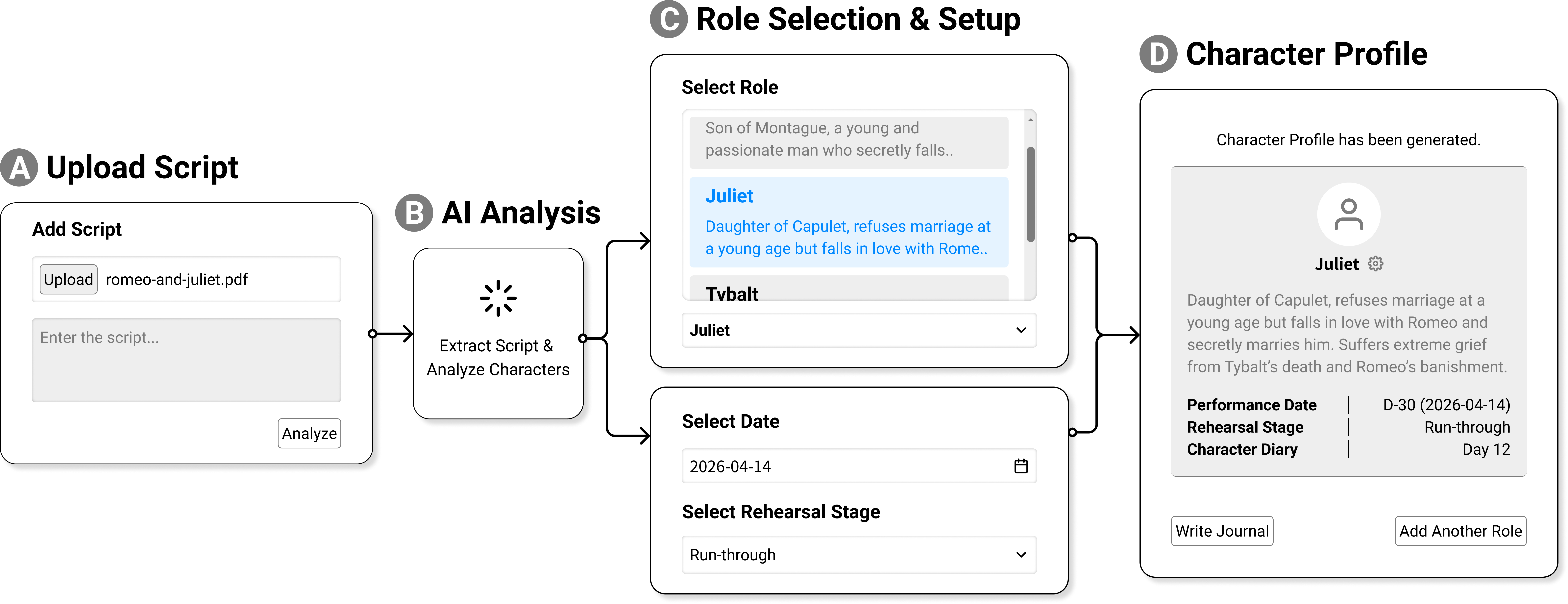}
  \caption{Onboarding flow, from script upload and role selection to auto-generated character profile and dashboard access.}
  \Description{Four-step onboarding flow diagram arranged left to right. Four labeled boxes are connected by arrows. The first box shows a file-upload icon. The second indicates an automated analysis step. The third shows selection fields for choosing a role, a date, and a stage. The fourth displays a character-profile panel with a button. Arrows connect each step linearly from left to right.}
  \label{fig:onboarding}
\end{figure*}

\begin{figure*}[t]
  \centering
  \includegraphics[width=\textwidth]{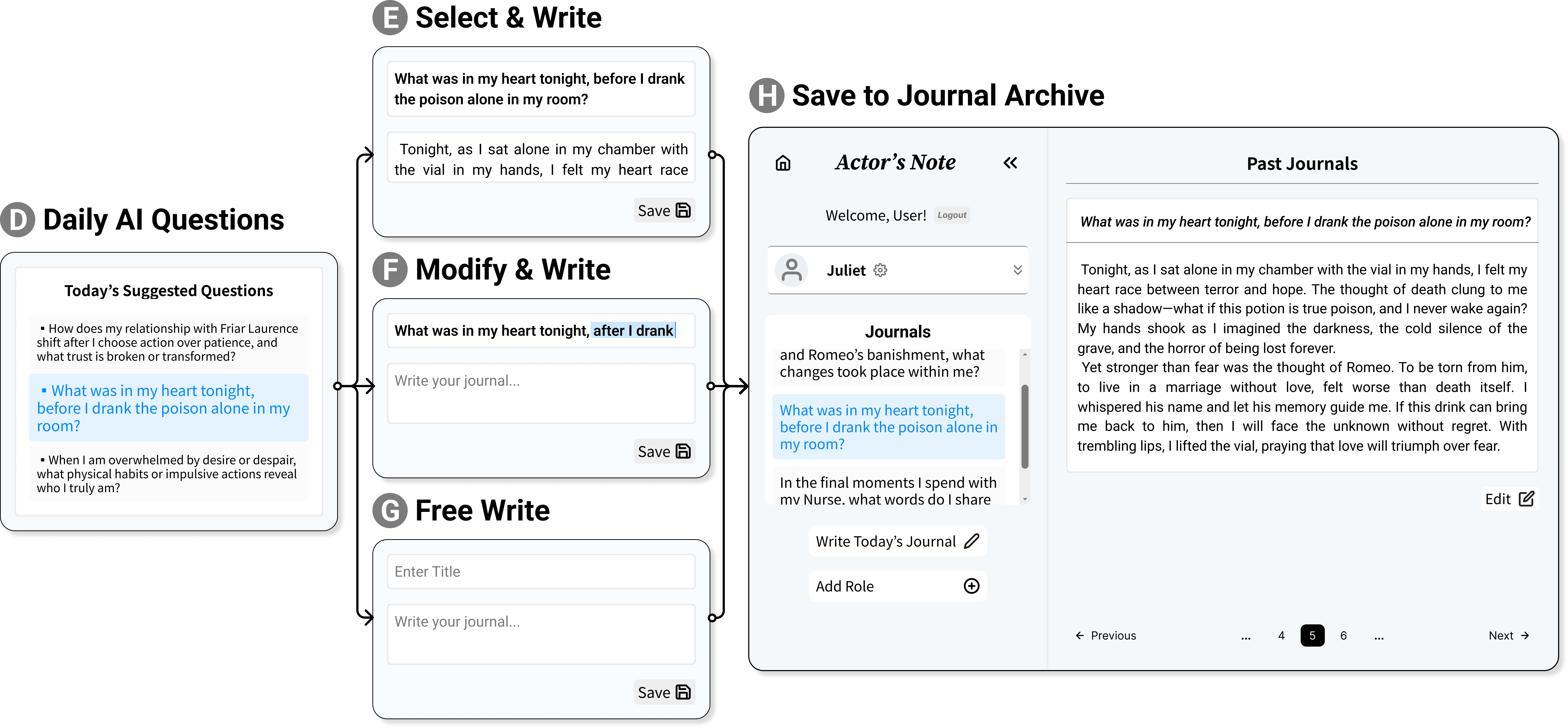}
  \caption{Daily flow, from receiving stage-aware questions to writing, saving, and archiving journal entries.}
  \Description{Daily-use branching flow diagram. A box at the top contains three suggested questions. Three arrows branch downward to boxes representing different writing options. All branches converge into a single box on the right labeled as a journal archive. The diagram forms a branched structure that reconnects at the end.}
  \label{fig:dailyflow}
\end{figure*}

\subsection{System Overview and Interaction Flow}
\textit{Actor's Note} is a web-based application designed to function as a journaling partner, augmenting actors' reflective practice. The system takes three inputs---script file \& role, performance date (D-Day), and current rehearsal stage---then uses an LLM to generate a character profile. For question generation, the prompt context includes the actor's role, a script summary, the rehearsal stage, and the performance schedule; the output appears as three question cards in the in-place journal editor. The actor selects a question, writes, and saves the entry. Past entries remain in an archive for review and revision. The interaction is organized into two phases:

\textbf{Onboarding flow.} At first use, the actor uploads a script (PDF). The system extracts the text, generates a summary, and identifies key characters. The user then selects their role, sets the D-Day and rehearsal stage, receives an auto-generated character profile, and proceeds to the dashboard (see Figure~\ref{fig:onboarding}).

\textbf{Daily flow.} The actor receives three stage-aware questions. They may use as-is, edit, or skip a topic, then write and save the entry. After writing and saving the entry, they return to the dashboard, where all previous journal entries remain accessible in an archive for reflection or editing (see Figure~\ref{fig:dailyflow}).

\subsection{Question Generation}
Question generation followed a constrained, maieutic schema: the model received five inputs---\{\textsc{script}, \textsc{role}, \textsc{stage}, \textsc{d\_day}, \textsc{character\_profile}\}. These fields jointly guide question semantics (e.g., focus and tone): \textsc{role} and \textsc{stage} primarily steer what aspect of the character to probe and when, while \textsc{script} provides contextual grounding. \textsc{d\_day} modulates temporal framing, and \textsc{character\_profile} helps maintain consistency in perspective and personality across sessions. The tool produced three questions per session, and each question was phrased in the second person (e.g., \q{What do you do when...?}) to support role immersion and character-centered reflection without generating content on the actor's behalf. The question-generation prompt was iteratively refined through a formative study with eight actors to ensure domain relevance and appropriate linguistic tone.

Questions were drawn from five themes of exploration: character concretization (daily habits or preferences), emotional exploration (feelings after key scenes), backstory completion (imagined past events), relationships \& change (interpersonal dynamics), and extreme scenarios (responses under stress). To maintain relevance, phrasing adapted to the rehearsal stage: early sessions emphasized grounding and motivation, while later sessions focused on emotional continuity and performance readiness. Representative questions for each theme appear in Table~\ref{tab:questions}.

\begin{table}[h]
  \renewcommand{\arraystretch}{0.5}
  \caption{Representative example questions generated under the five thematic categories.}
  \label{tab:questions}
  \centering
  \begin{tabular}{@{} l p{0.55\columnwidth} @{}}
    \toprule
    \textbf{Theme} & \textbf{Example Questions} \\
    \midrule
    Character Concretization & $\bullet$ \textit{What do you usually do when you are under stress?} \newline $\bullet$ \textit{What small habits reveal your personality?} \\
    \addlinespace
    Emotional Exploration & $\bullet$ \textit{How did you feel after discovering betrayal in scene X?} \newline $\bullet$ \textit{What do you think after a conflict with your closest friend?} \\
    \addlinespace
    Backstory Completion & $\bullet$ \textit{What do you remember about your life before the story began?} \newline $\bullet$ \textit{What do you regret most from your past, and how does it affect you now?} \\
    \addlinespace
    Relationships \& Change & $\bullet$ \textit{What do you feel toward someone you once trusted but now resent?} \newline $\bullet$ \textit{What do you notice changing in your bond with character Y?} \\
    \addlinespace
    Extreme Scenarios & $\bullet$ \textit{What will you do if someone humiliates you in public?} \newline $\bullet$ \textit{What would you choose if you got a chance to gain money through unfair means?} \\
    \bottomrule
  \end{tabular}
\end{table}

\subsection{System Implementation}
\textit{Actor's Note} is built as a web application using Next.js (React). PDF script files are parsed with PDF.js, and the extracted text is sent to the OpenAI API (GPT-4o-mini) to produce a script summary and a character list. Based on the selected role and rehearsal information, the system generates a character profile. For each daily entry, the system sends the role information (name and description), the rehearsal stage, the relevant script content, and a history of prior questions---to avoid duplication---to the LLM (GPT-4o). The model returns three candidate questions, which are displayed in the editor. Actors write directly in the editor and save their entries to the archive. Extracted scripts, generated profiles, questions, and journal entries are stored in Google Firebase Firestore.

%% file: section/5Method.tex
\subsection{Study Design}
This study employed a randomized crossover design with an initial baseline phase to examine the role of AI-generated questions in supporting actors' character journaling. Conducted in the wild, the study was embedded within participants' ongoing rehearsal processes rather than implemented in a controlled laboratory setting. We compared the AI-assisted condition with an unassisted freewriting condition—practice-as-usual baseline. The freewriting condition represented actors' typical journaling practice in rehearsal and served as the baseline. This baseline anchors the comparison in actors' natural journaling practice, enabling us to examine how the tool supports the continuity and enactment of character-journaling processes. Since character journaling is an introspective activity that may involve learning effects over time, the study design incorporated explicit checks for potential carryover effects. All study procedures—including journaling, surveys, and interviews—were conducted in Korean, as all participants were native Korean-speaking actors. Quotations cited in this paper were translated into English by a bilingual research team member for readability. The 14-day study consisted of three periods:

\begin{itemize}[noitemsep, topsep=2pt]
    \item \textbf{Period 1 (Days 1--2; Baseline):} All participants wrote character journals without AI assistance.
    \item \textbf{Period 2 (Days 3--8; First treatment):} Participants were randomly assigned to either the AI-assisted or unassisted journaling condition.
    \item \textbf{Period 3 (Days 9--14; Crossover):} Participants switched to the alternate condition.
\end{itemize}

\subsection{Participants}
Participants were recruited through online actor communities and public audition boards. The recruitment notice outlined the study's purpose, procedure, and eligibility criteria. To qualify, participants had to (1) be at least 19 years old and actively rehearsing a role during the study period (e.g., for stage or screen), and (2) have access to the study system on a mobile or desktop device.

Participants' ages ranged from 20 to 49 years ($M = 31.4$), and acting experience varied from 1 to 15 years ($M = 6.9$). Sixteen participants (55.2\%) primarily worked in theatre/musical, five (17.2\%) in film/TV, and eight (27.6\%) across both domains. A pre-study survey further indicated that 22 participants (75.9\%) had prior experience writing a character journal. While summary statistics are reported here, full participant-level demographic information is provided in \hyperref[appendix:B]{Appendix B} (see Table~\ref{tab:participant_demographics}).

Participants were randomly assigned to one of two treatment sequences. We refer to the two crossover sequences as Early-AI (AI-assisted-first sequence) and Late-AI (unassisted-first sequence).

\begin{itemize}[leftmargin=*, noitemsep, topsep=2pt]
    \item \textbf{AI-assisted-first sequence ($n = 14$):} Baseline session $\rightarrow$ AI-assisted character journaling (with system-generated questions) $\rightarrow$ Unassisted character journaling (no system-generated questions)
    \item \textbf{Unassisted-first sequence ($n = 15$):} Baseline session $\rightarrow$ Unassisted character journaling $\rightarrow$ AI-assisted character journaling
\end{itemize}

A total of 31 actors initially enrolled in the study. Two were excluded due to protocol violations and improper tool use, leaving 29 participants in the final analysis. All participants provided informed consent and received compensation equivalent to 36 USD.

\subsection{Data Collection}
We collected three types of data: self-reported surveys, system logs, and semi-structured interviews. 

\subsubsection{Surveys}
Participants completed a set of questionnaires to assess key process-level changes in reflective engagement and experience during journaling, using a 7-point Likert scale (1 = \textit{Not at all true}, 7 = \textit{Exactly true}).

\begin{itemize}[leftmargin=*, noitemsep, topsep=2pt]
    \item \textbf{Acting Confidence (AC):} Measured with three items adapted from the General Self-Efficacy Scale (GSE)~\cite{schwarzer1995generalized}. Example: \q{Through this character journaling activity, I developed confidence that I can successfully perform any scene of the character.} The scale showed strong internal consistency ($N = 93$, 3 items; Cronbach's $\alpha = .820$).
    \item \textbf{Cognitive Burden (CB):} Derived from three subscales of the NASA-TLX~\cite{hart1988development}: Mental Demand (\q{How mentally demanding was the process of writing this character journal?}), Effort (\q{How much effort did you have to put in to understand the character and write the journal?}), and Temporal Demand (\q{How much time pressure did you feel while writing the character journal?}).
    \item \textbf{Intrinsic Motivation (IM):} Six items adapted from the Intrinsic Motivation Inventory~\cite{ryan1982control} were used, covering three subscales: Enjoyment/Interest (e.g., \q{Writing the character journal was fun and enjoyable}), Perceived Choice (e.g., \q{I felt I could decide the content and direction of the character journal as I wished}), and Pressure/Tension (e.g., \q{I felt pressured to write the character journal}).
    \item \textbf{Character Understanding (CU) and Identification (CID):} A modified six-item version of the Character Identification Scale~\cite{cohen2001defining}. Items were split into two constructs:
    \begin{itemize}[noitemsep]
        \item \textbf{CU (Understanding):} e.g., \q{I have a clear understanding of the character's inner thoughts and feelings.}
        \item \textbf{CID (Identification):} e.g., \q{While writing the journal, I felt as if I became the character.}
    \end{itemize}
    \item \textbf{Narrative Transportation (NT):} Adapted from the Transportation Scale–Short Form~\cite{appel2015transportation}, modified for the journaling context. Example: \q{While writing the character journal, I was mentally immersed in the character's world.}
\end{itemize}

Additionally, a post-study survey measured usability and tool dependence. Usability was assessed using the System Usability Scale (SUS)~\cite{brooke1996sus} and the Perceived Usefulness scale~\cite{davis1989perceived} for AI questions. Dependence was measured with items adapted from the LLM-D12~\cite{yankouskaya2025llm}. Participants were also asked about future usage intentions and the likelihood of recommending the tool.

\subsubsection{System Log Data}
The system automatically logged every diary-writing session. Each record included pseudonymized session and participant IDs; writing date; the three AI-generated questions presented; the question selected and whether it was edited; the final submitted text; the delay from opening the editor to first keystroke; and the total writing duration. Final diary texts were retained for downstream text analysis. Sessions were labeled post hoc as AI-assisted when questions were presented and Unassisted otherwise. These logs were used to compare engagement and writing behaviors across conditions.

\subsubsection{Interviews}
At the end of the study, we conducted semi-structured interviews with participants via Zoom. Each interview lasted 30--40 minutes. The interviews were conducted individually, with two researchers participating: one leading the interview while the other assisted and took in-situ notes.

\subsubsection{Data Protection \& Ethics}
All procedures were approved by the institutional review board of the university where this study was hosted. All collected data were pseudonymized, and participation proceeded with informed consent.

\subsection{Analysis}
We conducted both quantitative and qualitative analyses to evaluate how the AI-assisted character journaling tool influenced actors' training experience.

\subsubsection{Quantitative Analysis}
\mbox{}\\
\textbf{Survey data.} 
We conducted our analysis using GLM with a randomized crossover design. The analysis followed a conditional analysis plan based on diagnosing carryover effects. First, carryover effects were diagnosed for each factor, and based on the diagnostic results, two different analysis methods were applied. Factors AC, CB, and IM used only data from period 2, which was unaffected by carryover. Specifically, an analysis of covariance (ANCOVA) was performed, controlling for baseline (Time-1) values as covariates. In contrast, for factors showing no carryover effect—CU, CID, and NT—we estimated the effect by combining data from both experimental periods (Period 2 and 3). We calculated the integrated estimates using a fixed-effect meta-analysis, which demonstrated superior model fit with lower AIC values (71.0--82.6 vs. 141--162) compared to repeated-measures ANOVA. Prior to analysis, basic statistical assumptions—normality, homoscedasticity, and linearity—were verified to ensure model validity (see Tables~\ref{tab:normality_check} and \ref{tab:linearity_check}, \hyperref[appendix:C]{Appendix C}).

\textbf{System Log Data.} Log data from all participants was consolidated into a single dataset. Time values were converted from milliseconds to seconds. Writing duration and writing start delay were winsorized at the top 1\% to reduce the influence of extreme outliers. Each entry was labeled as either AI-assisted or Unassisted.

\textbf{Text Processing.} We tokenized journal entries using the MeCab-ko morphological analyzer, restricting analysis to semantically meaningful parts of speech (common nouns, proper nouns, verbs, and adjectives).

\begin{itemize}[noitemsep, topsep=2pt, leftmargin=*]
    \item \textbf{Lexical Diversity:} Vocabulary diversity was measured with Herdan's C, a length-adjusted index.
    \item \textbf{Self-focused Language:} We defined two categories of self-referential language: (1) first-person pronouns (e.g., I, my, we), and (2) introspective verbs (e.g., think, reflect, feel).
    \item \textbf{Emotional Language:} Positive and negative sentiment words were identified using the KNU Sentiment Lexicon. We calculated both raw frequencies and normalized frequencies per 100 words.
    \item \textbf{Sentence Structure:} Sentences were segmented using regular expressions. We computed the number of sentences, mean sentence length (in words), and paragraph count (based on consecutive line breaks).
\end{itemize}

\textbf{Statistical Procedures.} Between-condition differences were tested using Welch's t-tests. Effect sizes were reported as Cohen's $d$ with 95\% confidence intervals estimated by 3,000 bootstrap resamples. Multiple comparisons were corrected using the Benjamini--Hochberg false discovery rate (FDR). In the AI-assisted condition, topic editing was coded as a binary variable, and editing rates with confidence intervals were estimated.

\subsubsection{Qualitative Analysis}
For the qualitative analysis, we analyzed the interview transcripts (recorded via Zoom, automatically transcribed with Fathom, and corrected manually) together with in-situ notes. Transcripts were coded in OpenQDA by four researchers following an open coding process. The finalized codebook comprised four top-level themes with corresponding subthemes:

\begin{itemize}[noitemsep, topsep=2pt, leftmargin=*]
    \item \textbf{Changes in Diary Experience:} writing process (immersion, thought expansion, confirmation, style shifts, enjoyment, prior habits/other); entry barriers (ease of starting, initial confusion); after AI removal (autonomy, absence, sustained influence).
    \item \textbf{Character Exploration \& Immersion:} memorable questions (emotion/state, relationships, motivations, new perspectives, affirmation); Facilitation (cognitive expansion/confirmation, internalization/identification, building emotion/relation/backstory).
    \item \textbf{Relationship with AI:} perceived role (tool; collaborator/partner; neutral; trust cues); comparison with human collaborators (pros/cons); creative agency (idea stimulation, confirmation, disruption).
    \item \textbf{Usability \& Design Suggestions:} usability issues (positive/negative feedback, UI/functional limits); improvement/extension ideas (feature requests, other ideas); practical applicability (best rehearsal stage, target users, limits); recommendation intent (positive, negative/conditional).
\end{itemize}

%% file: section/6Results.tex
This section presents both quantitative and qualitative results from the study. We begin by reporting overall usage patterns of \textit{Actor's Note} to contextualize participants' engagement with the tool. We then describe the survey and log results. Finally, we present qualitative findings from interview transcripts, highlighting how actors perceived the tool's role in their creative process and its potential applications in actor training. Each participant is referred to by a participant number (e.g., P01).

\begin{table*}[t]
  \caption{Period-2 GLM for carryover-affected factors and period-specific meta-analysis for non-carryover factors.}
  \label{tab:conditional_results}
  \centering
  \begin{tabular*}{\textwidth}{@{\extracolsep{\fill}} l l c c c c}
    \toprule
    \textbf{Factor} & \textbf{Analysis Method} & \textbf{$\beta \pm SE$} & \textbf{Statistic} & \textbf{\textit{p} (FDR)} & \textbf{Effect Size} \\
    \midrule
    \textbf{Acting Confidence} & Period-2 GLM & $\mathbf{0.992 \pm 0.264}$ & $t(26)=3.76$ & $\mathbf{<.001 (.003)}$ & $\eta_{p}^{2}=.352$ \\
    \textbf{Cognitive Burden} & Period-2 GLM & $\mathbf{-1.260 \pm 0.315}$ & $t(26)=-4.00$ & $\mathbf{<.001 (.003)}$ & $\eta_{p}^{2}=.381$ \\
    \textbf{Intrinsic Motivation} & Period-2 GLM & $\mathbf{0.602 \pm 0.235}$ & $t(26)=2.56$ & $\mathbf{.017 (.033)}$ & $\eta_{p}^{2}=.201$ \\
    Character Understanding & Period-Specific Meta & $0.241 \pm 0.192$ & $z=1.26$ & $.209 (.251)$ & $d=.198^{\dagger}$ \\
    Character Identification & Period-Specific Meta & $0.172 \pm 0.244$ & $z=0.70$ & $.482 (.482)$ & $d=.150^{\dagger}$ \\
    Narrative Transportation & Period-Specific Meta & $0.355 \pm 0.217$ & $z=1.64$ & $.102 (.152)$ & $d=.343^{\dagger}$ \\
    \bottomrule
    \multicolumn{6}{l}{\footnotesize \textit{Note:} Larger effects in bold. p-value codes—$^{\dagger} .050 < p < .100, ^{*} p < .050, ^{**} p < .010, ^{***} p < .001$; FDR-adjusted p in parentheses.}
  \end{tabular*}
\end{table*}

\subsection{Overview of Actor's Note Usage}
A total of 29 participants produced 371 valid diary entries during the study. Of these, 159 entries (42.9\%) were produced under the AI-assisted condition, and 212 entries (57.1\%) under the freewriting condition. Altogether, the dataset comprises approximately 162,000 characters (40,233 words) with a cumulative writing duration of 42.6 hours. On average, participants produced 12.8 entries each, though individual writing volumes varied. These figures demonstrate that \textit{Actor's Note} was consistently adopted throughout the study, providing a robust dataset for subsequent quantitative and qualitative analyses. Within the AI-assisted condition, 26 of 159 entries (16.4\%, 95\% CI: 11.4--22.9\%) involved topic edits. 12 participants (41.4\%) modified at least one question, though most entries were written without changes. Editing behavior varied across individuals, from frequent modification to none.

\subsection{Effects of AI-assisted Journaling}
Given carryover diagnostics, we report immediate between-condition effects with baseline-adjusted Period-2 GLMs for outcomes showing carryover (AC, CB, IM), and period-specific meta-analyses for outcomes without carryover (CU, CID, NT).

\subsubsection{What changed when AI stepped in}
AI assistance was associated with a substantial reduction in Cognitive Burden (CB) and meaningful gains in Acting Confidence (AC) and Intrinsic Motivation (IM). CB decreased with a large effect ($\beta=-1.260, q=.0026, \eta_{p}^{2}=.381$). AC increased ($\beta=0.992, q=.0026, \eta_{p}^{2}=.352$), and IM also rose ($\beta=0.602, q=.0331, \eta_{p}^{2}=.201$). In practical terms, daily, structured questions acted as a scaffold: they removed the \q{blank page} barrier, focused attention on character-centric reflection, and supported momentum once writing began.

\subsubsection{What did not change}
Factors related to acting competency showed no significant differences between conditions; within the study window, we did not detect a decrease under AI assistance. Character Understanding (CU) and Character Identification (CID) showed no condition differences, suggesting that core interpretive processes were not disrupted by assistance. Narrative Transportation (NT) exhibited a higher tendency under AI but did not reach significance at the group level (period-specific meta: $\beta=0.355, q=.152$) within the study window.

\subsubsection{Behavioral \& Linguistic Markers}
Alongside the survey results, we analyzed 371 valid log entries (159 AI-assisted, 212 freewriting) to examine how AI questions influenced the writing pattern and outcomes (see Table~\ref{tab:linguistic}).

\begin{table}[H]
  \caption{Comparison of Linguistic and Emotional Indicators Across Conditions. $\Delta$ = AI-assisted - Unassisted;}
  \label{tab:linguistic}
  \setlength{\tabcolsep}{3pt} 
  \centering
  \begin{tabular}{l c c c c c}
    \toprule
    \textbf{Measure} & \textbf{$\Delta$} & \textbf{$D$} & \textbf{95\% CI} & \textbf{$q$} & \textbf{Dir.} \\
    \midrule
    1st-person pronouns & +.015 & +.498 & (.30, .72) & $.001^{\star}$ & $\uparrow$ \\
    Negative emotion words & +1.09 & +.506 & (.31, .72) & $.001^{\star}$ & $\uparrow$ \\
    Self-reference ratio & +.011 & +.274 & (.08, .50) & $.020^{\star}$ & $\uparrow$ \\
    Positive emotion words & +.560 & +.260 & (.05, .48) & $.025^{\star}$ & $\uparrow$ \\
    Lexical diversity & +.028 & +.224 & (.09, .32) & $.025^{\star}$ & $\uparrow$ \\
    Sentence count & $-1.81$ & $-.211$ & $(-.42, .00)$ & .056 & $\downarrow$ \\
    Mean sent. length & $-1.69$ & $-.064$ & $(-.19, -.19)$ & .581 & -- \\
    Character count & $-5.40$ & $-.019$ & $(-.22, -.18)$ & .897 & -- \\
    Word count & $-0.95$ & $-.013$ & $(-.21, .19)$ & .897 & -- \\
    \bottomrule
    \multicolumn{6}{p{0.95\columnwidth}}{
      \scriptsize \textit{Note.} $^{\star} q < .05$ (Benjamini-Hochberg). Dir: Direction ($\uparrow$: AI > Unassisted). Lexical diversity: Herdan's $C$. Emotion/Pronoun counts are normalized per 100 words.
    }
  \end{tabular}
\end{table}

\textbf{Writing pattern.} Participants began writing faster under the AI-assisted condition ($M=141.7s$ vs. $227.7s$), but there was no significant difference between conditions in total writing time or length metrics.

\textbf{Linguistic and emotional expression.} AI-assisted entries demonstrated greater lexical diversity ($q \approx .025$) and more self-referential language (+1.5\%p; +38.5\% relative; $q<.001$). Overall self-referential word usage was also significantly higher ($q \approx .02$). Emotional expression increased as well, with both positive and negative words appearing more frequently, including a notable rise in negative emotion terms ($q<.001$).

\textbf{Sentence structure.} Freewriting entries contained slightly more sentences, though this effect was only marginal ($q \approx .056$). Other sentence-structure metrics showed no significant differences.

Taken together, the log data shows that AI assistance reduced cognitive load at the start of writing while fostering more diverse, self-reflective, and emotionally rich expression.

\subsubsection{Temporal Dynamics by Introduction Timing}
\mbox{}\\
\textbf{Interaction Effects Over Time.} Group $\times$ Time interactions were significant for CB ($F(2,54)=7.15, p=.002$) and marginally significant for NT ($F(2,54)=2.45, p=.096$). Among the results with negligible interaction effects, a significant main effect of time was observed for CID ($F(2,54)=5.24, p=.008$) (see Table~\ref{tab:anova_results}).

\begin{figure}[h]
  \centering
  \includegraphics[width=\linewidth]{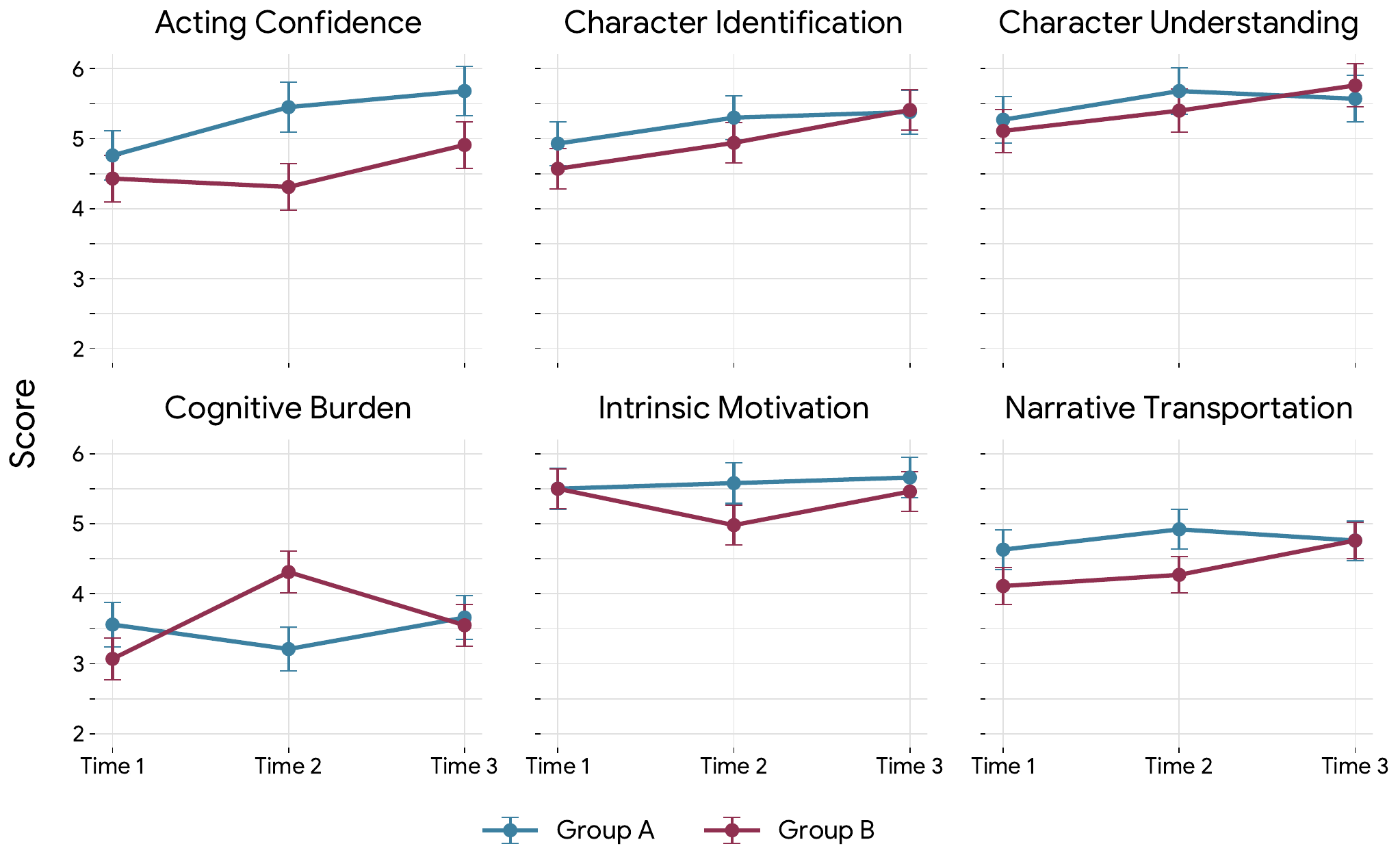}
  \caption{Temporal trajectories by group. Scores increased during AI phases (except Cognitive Burden), showing comparable magnitudes across groups.}
  \Description{Six line charts showing two labeled groups across three timepoints. Each panel has a horizontal axis labeled Time 1, Time 2, and Time 3, and a vertical axis labeled Score. Two lines appear in every chart, distinguished by different marker shapes and identified in the legend as Group A and Group B. Acting Confidence: both groups increase over the three timepoints, with Group A slightly higher. Character Identification: both groups rise gradually and stay close together. Character Understanding: both groups increase in parallel. Cognitive Burden: Group A dips at Time 2; Group B peaks at Time 2; both return toward similar levels at Time 3. Intrinsic Motivation: both groups dip at Time 2 and rise again at Time 3. Narrative Transportation: both groups remain close, with small changes over time.}
  \label{fig:temporal}
\end{figure}

Trajectory shapes differed by treatment timing. In the Early-AI group, CB dropped sharply by Time-2 and then ticked up slightly by Time-3, while AC and NT rose through Time-2 and stabilized. In the Late-AI group, the AI-free interval (T1$\to$T2) was marked by rising CB and temporary dips in AC and IM; after AI was introduced (T2$\to$T3), CB fell steeply, and AC/IM recovered. NT followed the same recovery pattern post-treatment. These temporal profiles align with the carryover pattern diagnosed earlier (see Figure~\ref{fig:temporal}).

\textbf{Pairwise Post-hoc test.} Tukey-adjusted pairwise post-hoc comparisons---applied to CB, AC, and NT, the factors that showed medium-or-larger effects---clarify where trajectories diverged across groups and times.

For CB, the only between-group difference emerged at Time 2, with Early-AI < Late-AI ($p=.006$). Within groups, Late-AI showed a rise from Time 1 to Time 2 ($p=.0006$) followed by a drop from Time 2 to Time 3 ($p=.0561$), whereas Early-AI exhibited no significant within-group contrasts. For AC, Early-AI > Late-AI at Time 2 ($p=.0049$) and Time 3 ($p=.0475$). Within Early-AI, confidence increased relative to baseline at Time 2 ($p=.0371$) and Time 3 ($p=.0032$). For NT, the between-group contrast favored Early-AI at Time 2 at a trend level ($p=.0774$) and was non-significant at Time 1 and Time 3. Within Early-AI, there were no significant time contrasts, whereas Late-AI showed a significant increase from Time 1 to Time 3, $p=.0128$ (see Table~\ref{tab:pairwise_comparisons} in \hyperref[appendix:C]{Appendix C})

\subsubsection{Usability and User Perceptions (Post-Study Survey)}
\begin{figure*}[t]
  \centering
  \begin{subfigure}[b]{0.38\textwidth}
    \centering
    \includegraphics[width=\linewidth]{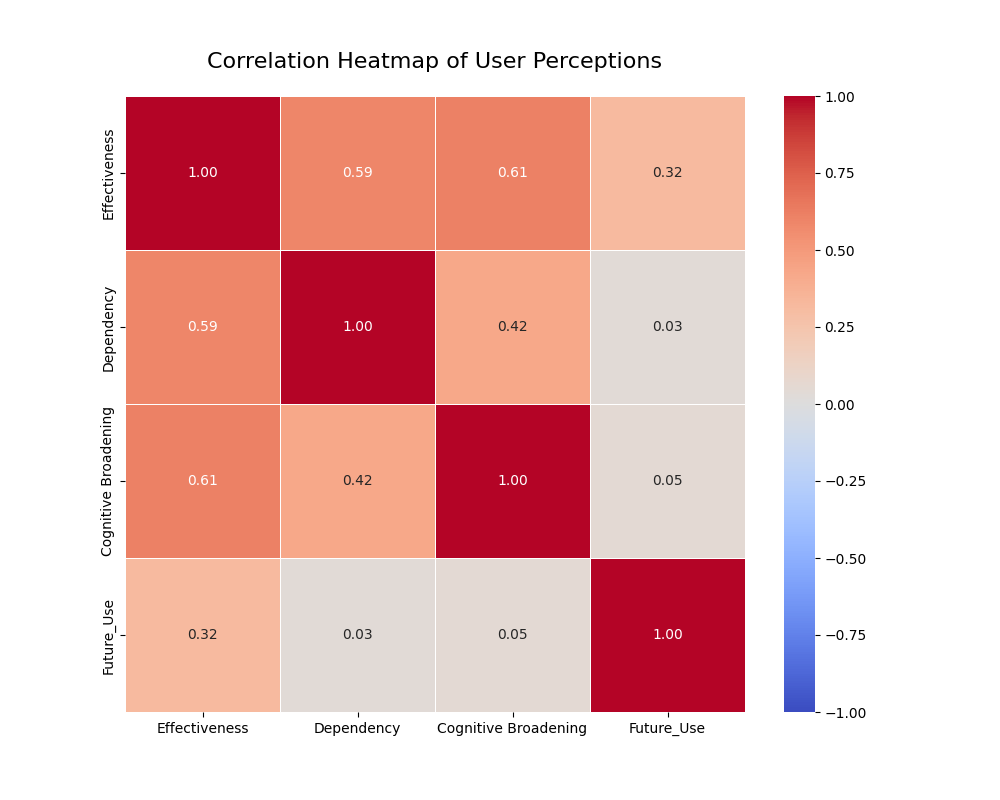}
    \caption{Correlation Heatmap}
    \label{fig:poststudy-heatmap}
    \Description{A 4×4 correlation matrix labeled with “Effectiveness,” “Dependency,” “Cognitive Broadening,” and “Future Use” along both the rows and columns. Each cell contains a numeric correlation value. Diagonal cells show 1.00. Off-diagonal values range from approximately 0.30 to 0.60. Cells with higher values appear visually darker while lower values appear lighter. A vertical scale bar appears on the right with numeric values from –1.0 to 1.0.}
  \end{subfigure}
  \hfill 
  \begin{subfigure}[b]{0.58\textwidth}
    \centering
    \includegraphics[width=\linewidth]{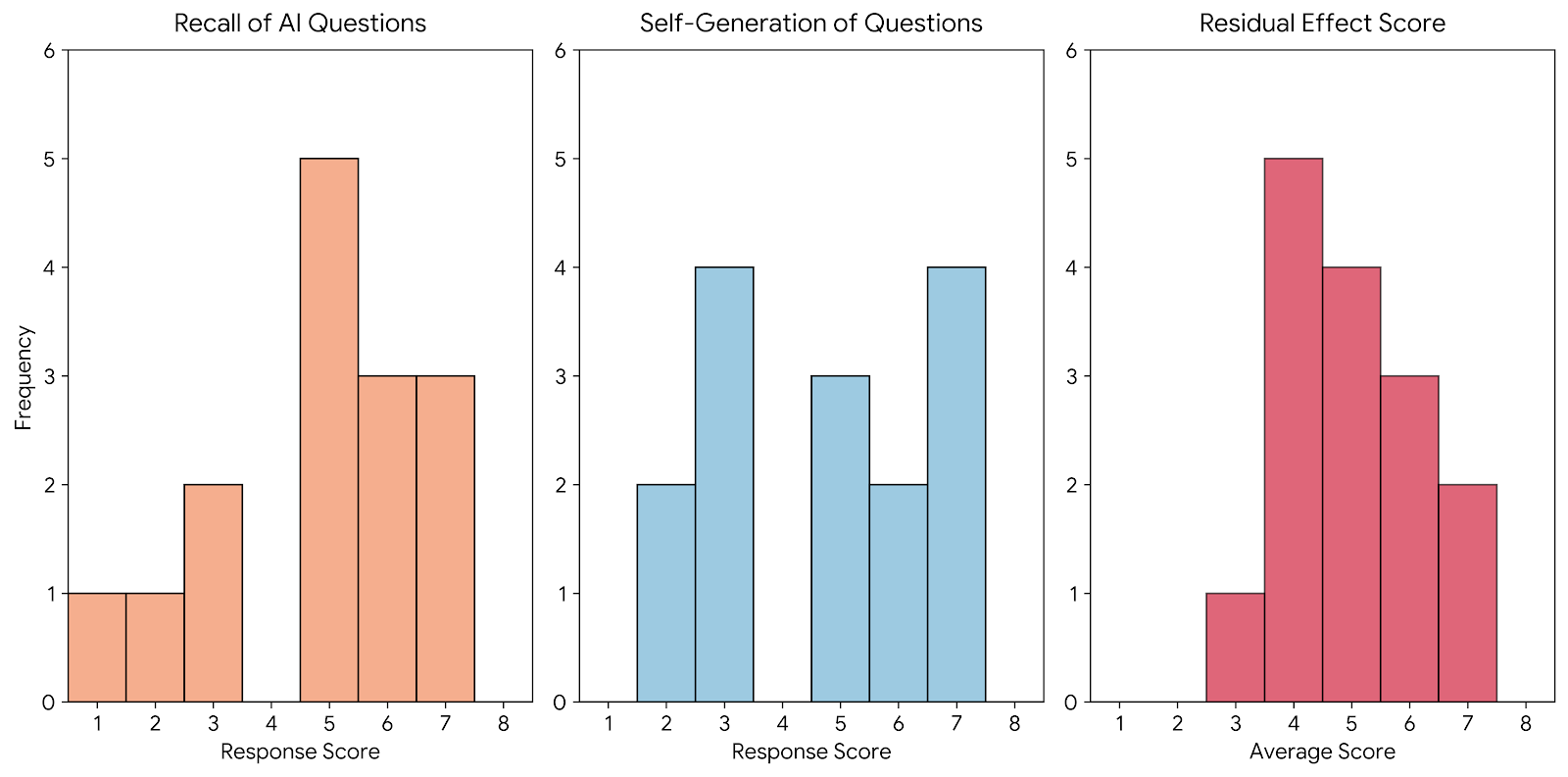}
    \caption{Residual Effect Histograms for Early-AI group}
    \label{fig:poststudy-histograms}
    \Description{Three histograms displayed side by side. Each histogram includes a horizontal numeric scale representing response scores and a vertical scale labeled “Frequency.” Left histogram titled “Recall of AI Questions”: Bars appear across the lower and mid score range with one taller bar near the center. Middle histogram titled “Self-Generation of Questions”: Bars cluster in the mid-to-higher score range with several tall bars. Right histogram titled “Residual Effect Score”: Bars span the full score range with relatively even heights and no single dominant peak.}
  \end{subfigure}
  
  \caption{Post-study survey results showing (a) correlations between user perception metrics and (b) residual effects of AI usage on recall and self-generation. Histograms indicate a positive lasting impact of the tool.}
  \label{fig:poststudy-results}
\end{figure*}

Overall usability was high: the SUS score was 82.76, commonly interpreted as Excellent. To probe user perceptions, we examined correlations among four composites: Perceived Effectiveness \& Utility, AI Tool Dependency, Cognitive Broadening, and Future Use. Perceived Effectiveness \& Utility correlated strongly with both Cognitive Broadening ($r=.61$) and AI Tool Dependency ($r=.59$); Tool Dependency correlated moderately with Broadening ($r=.42$). Future Use showed weak or negligible associations with the other composites, suggesting that continued adoption may depend on situational or organizational factors beyond immediate efficacy (e.g., rehearsal schedules or director preferences; see Figure~\ref{fig:poststudy-results}a).

We also observed descriptive residual effects among Early-AI participants. They exhibited residual effects in their experience with AI questions suggestion: the tendency to recall AI-questions later ($M=5.00$) and to generate similar questions independently ($M=4.73$), with the mean for both questions being 4.87 (see Figure~\ref{fig:poststudy-results}b).

\subsection{Catalyzing Actor Journaling: Initiation, Depth, and Transfer}
For actors, character journaling was an important but daunting task. The primary barrier was not knowing what to write when facing a blank page. \textit{Actor's Note} addressed this by easing initial intimidation and breaking creative inertia. The following sections present interview findings, supplemented by quantitative evidence.

\subsubsection{Initiation \& Agency: Overcoming the Blank Page}
\mbox{}\\
\textbf{Reduced cognitive load through directional cues.} Actors reported that, when writing journals without AI, they often faced a vague uncertainty---\q{What should I write today?} However, once \textit{Actor's note} began suggesting daily questions, they described the experience as \q{like a ship sailing toward a set destination,} giving their writing direction and easing their minds. As P22 noted, \q{When writing alone, I'd think, `Am I wasting my time?' and I'd want to quit. With this tool, it felt like sailing toward a set destination.} They credited the AI with guiding them to sustain their writing through to the end. P29 shared, \q{I tended to lump things together broadly, but I was able to address specific details with Actor's note,} indicating that the specific questions posed by the AI translated initial overwhelm into structured goals, providing new momentum for writing. System logs showed that after introducing AI, participants' average time to start writing decreased, and survey results likewise showed a significant reduction in CB under the AI-assisted condition.

\textbf{Securing initiative through offering choices.} \textit{Actor's Note} supported user autonomy by offering three questions rather than a single one, allowing actors to choose freely and even revise the suggested questions. P7 remarked, \q{If there had been only one question, it would have felt forced, but having options made it less burdensome.} Furthermore, some actors used all three questions, writing multiple entries per day: \q{Instead of writing just one entry a day, since I had three, I wrote all three.} (P22). As this quote suggests, diverse questions spurred additional writing. Furthermore, rather than passively accepting the AI's questions, users actively explored topics that piqued their interest; for instance, if they didn't like a particular question, they refreshed the page to get a different one. As P5 noted, \q{When I hit refresh, the topic kept changing \dots I refreshed a few times until the desired topic appeared,} indicating that actors actively steered the system to seek inspiration beyond the AI's initial options. Consistent with this autonomy-supportive pattern, survey responses indicated higher Intrinsic Motivation (IM) under AI-assisted conditions.

\subsubsection{Deepening Immersion and Writing Experience}
Once writing began, the AI's questions guided actors into deeper reflection, enhancing immersion. Many participants reported writing longer and with greater intensity than they expected when with AI. This was because AI consistently stimulated their thinking by posing varied questions. For instance, P11 recalled, \q{When asked about a specific emotion, I suddenly found myself writing a lot. It felt like everything came flooding out at once.} We observed multiple instances in which a single, specific question evoked memories and emotions, yielding unexpectedly deep, rich content. Some even mentioned feeling \q{a kind of euphoria} while writing, suggesting a heightened state of immersion associated with the AI's involvement. Text analysis revealed that lexical diversity increased significantly under AI assistance, indicating linguistically richer writing. Furthermore, NT showed a small positive effect in the AI condition.

\textbf{Motivation from Perceived Intelligence.} Several participants built their trust after seeing the tool summarize the play and map character relationships based on the script, recognizing it as an `intelligent being' that understood the play. P7 mentioned \q{I wondered if it could grasp such a complex play at first, but seeing how well it pinpointed the plot and character relationships made me trust it more.} Multiple participants also shared that their attitude shifted after a few days when they felt the tool posed \q{sharper questions than expected,} tailored to their writing style (P11, P12, P23). P23 specifically explained, \q{At first, I wondered, `Can this really analyze the script properly?' Later, when the AI suggested dark questions, I gained trust, realizing, `It's analyzing what I wrote to give me those questions.' I wrote every single day without fail.} This perception---\q{the AI understands my writing}---became a powerful motivator, serving as the decisive factor that led the participant to \q{write in the diary without missing a single day.} In other words, belief in the AI's perceived intelligence sustained long-term interaction and immersion.

\subsubsection{A Questioning Framework That Persisted After AI}
The impact of \textit{Actor's Note} persisted even after the treatment ended. Participants who had written journals with the tool for six days distinctly noticed the void once the AI feature was removed. Several reported feeling a bit lost when the questions suddenly disappeared, re-experiencing the initial bewilderment when journaling alone without AI.

P25 noted, \q{This was something I used to do by myself, but when it suddenly disappeared after receiving help, I thought, `What should I do?'} This revealed a dependency on AI assistance. It ironically underscores how effective AI's \q{providing a starting point} function was for users and how significant a psychological support it offered.

Furthermore, we observed that actors internalized how the AI posed questions. Even when writing alone without AI, they recalled and reflected on the AI's previous questions. P29 explained, \q{It wasn't something that the specific questions remained; rather, the way of questioning did---such as thinking in the opposite direction.} This demonstrates that they learned not the content from the AI, but its framework for thinking. P24 also mentioned that right after the experiment ended, while preparing for a new audition, they thought, \q{What questions would the AI have asked for this piece? Which aspects should I probe further to understand it?} This indicates that \textit{Actor's Note} functioned beyond a temporary idea catalyst, instilling a persistent questioning framework within actors' thought processes. In the post-study survey, the Early-AI group showed residual effects: a strong tendency to recall AI questions ($M=5.00$) and to generate similar questions themselves ($M=4.73$), even after assistance ceased. These results suggest that AI's influence became internalized within participants' thought processes.

\subsection{Driving Deeper Inquiry: Validation, Confidence, and Internalization}
As journaling progressed, \textit{Actor's Note} functioned as an engine for deepening and broadening actors' inquiry. The AI's structured questions guided participants to cognitively explore new territories while simultaneously validating existing thoughts and instilling confidence.

\subsubsection{Cognitive Extension Beyond the Script}
\textit{Actor's Note} encouraged actors to expand their thinking beyond what was explicitly written in the script. Its effectiveness was particularly pronounced in building backstories, such as a character's history or hidden facets. P22 noted, \q{I used to build backstories from what the script revealed, but the AI asked about unseen aspects, which helped me flesh out the character much more fully.} Consistent with these reports, linguistic analyses showed a significant increase in lexical diversity, which may reflect actors' exploration of previously unarticulated aspects of their roles. Questions that delved into the context behind actions also provided fresh inspiration for the actors. For instance, P27 gained insight after being asked, \q{How would this character behave when stressed?} She shared, \q{I had only thought of this character as a very happy person, but since everyone experiences stress, I realized I needed to consider that aspect. It made me think about aspects not explicitly shown in the script.} P29 was surprised by the AI's question, asking: \q{Expectation and burden? These are completely opposite emotions---why ask about them?} and mentioned the experience of imagining these opposing feelings simultaneously. She explained, \q{I'd never considered burdens for this character before, but asking about such opposite emotions made the character feel more three-dimensional.} In this way, the AI prompted actors to contemplate hidden facets and opposing sides of their characters, helping them build more multi-layered and realistic portrayals.

\subsubsection{Reinforcing Interpretive Confidence}
In addition to opening new avenues, AI reinforced actors' confidence in their existing interpretations. As P6 noted---\q{When questions similar to my own thoughts arose, I felt relief and certainty, thinking, `Ah, I wasn't wrong.'} P19 also remarked, \q{The AI pinpointed one of the most crucial aspects I focused on throughout the performance, helping me organize my thoughts and solidify my conviction.} In this way, the AI's questions became an opportunity for actors to reassemble scattered fragments of their thoughts, granting legitimacy and belief in their characters' actions. P21 mentioned confirming the legitimacy of their character's actions through journaling. She explained, \q{While journaling, I concluded that if the character had lived this kind of life and felt these emotions, it would make sense for them to act this way.} This illustrates how AI-generated questions can yield decisive understanding and empathy for a character's motives, instilling trust in the character. This cognitive certainty and psychological support translated into a statistically significant increase in AC under AI-assisted conditions, as confirmed by quantitative results.

\subsubsection{Deepening of Internalization and Identification}
This exploration process often fostered deeper empathy and identification with the character. \textit{Actor's Note} framed questions in the second person (e.g., \q{When you were in that situation \dots}), as if addressing the character, and actors responded in the first person. As a result, actors wrote journals as if embodying the character and experiencing situations from the character's viewpoint. As P22 noted, \q{The questions were phrased like, `How did \textbf{you} feel when \textbf{you} did this?'---so it felt like I was writing a diary as the character \dots.} This shift in subject perspective enabled the actors to internalize and genuinely feel the character's emotions and thoughts as their own. Through this process, actors moved beyond analysis to experience the character's inner world and form convictions about their actions and emotions. Although the CID between treatment conditions was not significant, linguistic indicators showed more self-referential expression under the AI-assisted condition: the proportion of first-person pronouns increased relative to the free condition, and the proportion of self-referential vocabulary also increased. In addition, the use of both positive and negative emotional words rose, suggesting richer and more honest inner exploration. This suggests the formative process---the \q{pathways of thought and linguistic traces} leading to identification---shifted.

\subsection{\texorpdfstring{\scalebox{0.93}{The Formation of a New Creative Partnership}}{The Formation of a New Creative Partnership}}

As interactions with AI continued, a unique creative partnership emerged between actors and \textit{Actor's Note}, moving beyond a simple user--tool relationship.

\subsubsection{From ``Tool'' to ``Collaborator''}
In the early stages, many participants regarded the AI as a tool; however, as interactions accumulated and the precision and contextual relevance of its questions became apparent, perceptions gradually shifted toward seeing it as a collaborator. P11 reflected: \q{At first I thought of it as a robot, but after 2--3 days it felt like a collaborator who shared in the outcomes.} Many participants described working with \textit{Actor's Note} as a new form of working---\q{alone but not alone.} P7 said, \q{At first, it was like someone just throwing out topics, but later it felt like someone checking in, a bit like a mother.} P22 also noted, \q{After the AI came along... it felt like a secret weapon,} indicating that AI added a companion-like element to the isolated, solitary task of character work. In sum, participants increasingly perceived it not merely as an auxiliary tool but as a creative partner, reflecting a shift toward wanting to work together.

\subsubsection{Complementary Roles with Human Colleagues}
Actors also discussed comparisons between collaborating with AI and with human partners. The most frequently cited advantages of AI were psychological safety and objectivity. P22 said, \q{When analyzing scripts with people, it can feel like you must argue and prevail logically, which is exhausting. With AI, there's no need for that, which was helpful.} Participants frequently reported that AI, asking from a third-party perspective without vested interests or emotional entanglement, enabled them to write more candidly without fearing others' reactions. P17 put it: \q{When working with people, I sometimes couldn't fully express my thoughts.} Several actors also found the AI's questions useful, likening them to feedback from an audience perspective. P1 mentioned, \q{An actor's performance is ultimately for the audience, so it was helpful that the AI's questions felt like they came from an audience perspective.} This suggests that AI can provide a generalized audience viewpoint that helps balance an actor's interpretation.

Meanwhile, participants also emphasized the creative benefits of working with human colleagues. Whereas AI collaboration can be less emotionally taxing, collaboration with people affords immediate back-and-forth that surfaces unexpected ideas. P3 remarked, \q{I noticed that AI could ask questions like what I received from directors. However, to reach a more creative feel, conversations with people are still better.} This suggests that AI and humans can play complementary roles in actor training. AI is effective as an objective coach, whereas human colleagues excel as intuitive, improvisational collaborators.

\subsubsection{How Actors Retain Agency with AI}
All participants agreed that AI facilitated---rather than impeded---creativity. They also described AI as a \q{stimulant for overcoming blocks} (P20) and a \q{booster} (P27), converging on its creativity-enhancing role. Yet a few expressed concerns. P1 noted, \q{When AI was introduced suddenly, I struggled to follow my usual way of journaling,} indicating that AI's interventions could temporarily disrupt creative flow. P6 remarked, \q{The AI kept asking depressing questions; even though I wanted to write about positive aspects, it felt as if it was pushing the mood in one direction.} Even when they encountered minor discomfort, participants took proactive steps to maintain creative agency. They reported, \q{So I just chose a different question} (P5) and \q{I skipped questions I didn't like} (P22). Under AI-assisted conditions, actors modified 16.4\% of diary entries, and 41.4\% of participants changed the topic at least once. Several emphasized that authorship remained theirs---\q{AI just tossed out topics, and since I was the one choosing and writing, I felt I was writing on my own} (P19); \q{I didn't feel like AI did anything huge. It just felt like exactly the kind of reference that helped me---and that was actually nice} (P8). Participants did not perceive their creativity as being suppressed by AI; rather, they adopted an agentic stance, treating AI as a creativity-support tool and using it selectively.

\subsection{Integrating Actor's Note into Rehearsal: When, How, and With Whom}
Finally, participants offered specific usability feedback on \textit{Actor's Note}, practical guidance on when and how to use it in rehearsal and training settings, and suggestions for feature enhancements.

\subsubsection{Timing \& Workflow Fit}
Most participants agreed that the tool was most effective during the initial phase of script analysis; using AI during reading or table work broadened interpretive possibilities. P27 said, \q{It seems more beneficial in the early stages because interpretations are already solidified later on.} Participants noted that using it in later stages---during filming or performance---could introduce confusion. P26 observed, \q{During filming or on stage, there is already an agreed-upon directorial plan; if I keep writing a journal, the direction might shift and cause confusion.}

Opinions diverged on the precise timing of AI introduction. Some preferred to first sketch character basics and then use AI's questions to fill gaps and enrich details---\q{It would have been more effective after I established a basic foundation myself and then use its questions} (P1). Conversely, several participants said prolonged solo writing reduced interest, so they favored earlier introduction to maintain momentum. P20, a Late-AI participant shared, \q{After about a week, I ran out of ideas and was about to stop writing---but then I received the AI questions, I managed to keep going.} Additionally, actors suggested sequencing the AI's questions from broad to specific for a more natural flow and, where possible, supporting a chronological writing order, as P9 said, \q{It would have been better if the initial questions provided an opportunity to write chronologically and then moved into the details.}

\subsubsection{Personalization, Collaboration \& Adoption}
Actors proposed features to enhance personalization and collaboration. The most frequently requested feature was a team-shared diary: \q{If the character journals each of us writes could be viewed by the director and other actors, it would be a huge help} (P29). Participants noted that team-wide sharing would help align interpretations and facilitate discussions across roles. Another key request concerned customizing question content and format; because actors explore different facets of their characters, they wanted to select questions directly or specify categories--\q{Categorize questions by scene} (P3), \q{Allow more questions when needed} (P15).

Several participants also wanted \textit{Actor's Note} to evolve beyond a question generator into a more interactive dialogue partner. As P9 put it, \q{It would be great if it could give feedback on my journal or pose follow-up questions based on them,} highlighting a wish for an AI that responds to their inputs. Participants expressed positive views about the tool's primary beneficiaries and its future adoption: many noted that the tool could serve as a structured exploration guide for theater/film students, novices, and actors who have relied more on intuition than paperwork. Above all, many reported intentions to reuse the tool and recommend it to colleagues, as P16 mentioned, \q{Having tried it, I found it effective, so I plan to use it again when preparing my next character.} These reactions suggest that \textit{Actor's Note} holds practical value for rehearsal and production contexts and indicate strong intentions to reuse it.

%% file: section/7Discussion.tex
\subsection{\texorpdfstring{\scalebox{0.93}{Journaling Is Effective. What’s Better with AI?}}{Journaling Is Effective. What’s Better with AI?}}

As discussed earlier, journaling itself substantively supports character exploration. In practice, however, sustained journaling is hindered by high cognitive load, time pressure, the ``blank page'' problem, and limited feedback. In our study context, all participants were required to journal over a set period; accordingly, Character Understanding (CU) and Character Identification (CID) increased over time. By the end of the study, many scores approached the upper range of their scales, and condition differences were not statistically pronounced—consistent with a plausible ceiling effect. This does not imply that AI assist failed to raise understanding or identification of their characters; rather, it suggests that the baseline utility of journaling is already strong and that sustained practice stabilized core interpretive competencies across participants.

Even so, the AI tool delivered clear process-level advantages. It reduced cognitive burden (CB), increased intrinsic motivation (IM), and boosted acting confidence (AC). We attribute these gains to topical suggestions and staged scaffolding, as well as to a partner-like interaction that lowered the entry barrier presented by the blank page and reduced ongoing maintenance costs. The core significance of the tool, therefore, is that it supports and streamlines a good journaling practice without undermining journaling’s inherent strengths, making the process lighter and more sustainable.

Participant feedback points to concrete avenues for improvement. Adaptive control of question difficulty and pacing, topic customization, actionable feedback, and interactive follow-up questions are likely to further enhance motivation, self-efficacy, and engagement and, in some contexts, may even augment acting competencies beyond manual journaling. With these refinements, the tool can improve the quality and sustainability of actors’ experience even in contexts where understanding and identification already sit near an upper bound.

\subsection{Designing AI as a Maieutic Partner. What Kind of AI Do Actors Want and Need?}
We frame the system—a creativity support tool for actors—as a maieutic partner that asks rather than writes: it elicits reflection through targeted, context-aware questions while keeping authorship with the actor. Character work—especially journaling in this study—is necessary yet solitary and difficult. Accordingly, the tool should help actors think, not think on their behalf, so that core capacities remain with the actor. Participants reported that this stance preserved agency, reduced start-up friction, and created a psychologically safe space for exploration.

\subsubsection{Why this stance works}
This stance works for several mechanistic reasons. Directional cues reduce cognitive burden and start latency by indicating where to begin, dissolving blank-page paralysis (§1.2.1). Context-aware prompts that are sensitive to script, role, and stage make the system feel legible and ``understanding,'' which sustains engagement over time (§§1.3.2, 1.5). Second-person prompts invite in-character, first-person responses, focusing attention on internal state and action; entries exhibit greater lexical diversity, more self-reference, and richer positive and negative affect (§1.2.3). Agency is maintained in practice because the system asks instead of writing and exposes lightweight controls—choose, refresh, edit, and skip (§§1.1, 1.5.3). To avoid over-reliance, we deliberately surfaced only three questions per turn and omitted prescriptive how-to scaffolds; participants reported retention of their own voice. Notably, many internalized the questioning frame and continued to self-generate similar prompts after assistance was removed (§1.3.3). These observations yield concrete design implications for a creativity tool for actors.

\subsubsection{Design Implications}\mbox{}\\
\textbf{\textit{Outputs should remain questions or directional cues}} rather than interpretations, with choices that let actors steer, so authorship remains firmly theirs; this was associated with a sense of guided progress without loss of ownership (P22, P19, P8).

\textbf{\textit{Trust should be built so the AI feels like a partner}}: competence should be shown early, and questions should be made legible and grounded by tying them to specific lines or beats or to themes surfaced in recent entries, including brief plot summaries or character-relationship maps when appropriate; several participants perceived the system as understanding their play and their work, which subsequently motivated longer-term engagement (P7, P23; also P11, P12, P23 regarding trusting, contesting, or redirecting questions).

\textbf{\textit{Agency should be protected}} through lightweight controls and customization, including multiple prompt options per turn, refresh/edit/skip, an optional no-prompt mode, direct selection and categorization, and the ability to request additional questions on demand (P5, P22, P3, P15).

\textbf{\textit{A balance between novelty and confirmation}} is important so that fresh angles that reframe beats are complemented by salient, expected points; participants associated this mix with expanded interpretation and reinforced core choices, contributing to confidence (P6, P19, P21, P22, P27, P29). Because productions are collaborative, optional sharing modes can bridge private reflection and ensemble needs by allowing individual writing that can be shared with teams or directors when helpful (P29).

\subsection{Timing Matters. When Should AI Step In?}
We examined how the timing of AI introduction shapes growth trajectories. Introducing AI early quickly reduces uncertainty and the blank-page burden, and introducing AI later deepens actor’s narrative transportation.

\subsubsection{What participants said \& what happened}
Participants’ accounts align with these patterns while differing in preference. Some preferred a later introduction, indicating that questions were more effective after they had first established a basic interpretation on their own and that the tool worked well for complementing missed elements after script analysis (P1, P17). Others favored an earlier introduction, noting that an extended solo period could reduce interest, and that early support helped maintain momentum when motivation flagged (P20, P22).

Sequence effects in the cross-over design corroborate these perspectives. In the Early-AI sequence, AC and IM rose rapidly while CB dropped immediately, and AC remained higher even after AI removal, exhibiting a fast-rise-then-stabilize echo. In the Late-AI sequence, the no-support period from T1 to T2 brought a spike in CB ($p < .001$), indicating heightened difficulty, but once AI was introduced from T2 to T3, NT increased significantly relative to baseline ($p < .05$), a pattern not observed in Early-AI. By the end of the cross-over (T3), outcomes largely converged, indicating that both approaches are effective; the question is not which is universally better, but which timing aligns with the actor’s needs. These findings motivate specific pedagogical guidance.

\subsubsection{Pedagogical Implications}
\textit{For novices or highly intuitive actors} who face substantial uncertainty at early stages, an early introduction provides starting points, direction, and a questioning frame that lowers cognitive burden and stabilizes confidence quickly. And \textit{for actors already comfortable with journaling,} a later introduction can be more effective; emphasizing counter-affect strategies, counter-arguments, reverse-questions, and scene or transcript expansion promotes deeper qualitative immersion. \textit{A hybrid approach is also viable}: introducing AI from the outset with very basic, high-level questions can help actors build their foundation and maintain momentum, and once that base is set, shifting toward the advanced can increase depth. When applying these strategies, one caution is worth noting—possibly especially for novices. Removal of AI may require a brief adaptation period, as suggested by interview evidence and downward tendencies in CU and NT. A plan for gradual independence is therefore advisable, involving a gradual reduction in the frequency and difficulty of questions while preserving user agency through selection, editing, and refreshing.

\subsection{Limitations \& Future Work}
We did not measure acting quality, as the study focused on reflective and cognitive processes rather than performance outcomes; journaling primarily supports deeper reflection rather than yielding immediate performance gains. The AI-generated scaffolding in our system functions as a coherent design, meaning the findings characterize its influence on reflective engagement without disentangling the effect of structured prompting from the effect of AI adaptivity/generation. Future work may incorporate additional non-AI scaffolding baselines (e.g., a static, curated question set) to parse these layers in a more controlled comparison. Overall, the results should be interpreted as reflecting the impact of adding AI-supported scaffolding relative to actors’ typical journaling practice.

Regarding timing constraints, we included a two-day baseline to re-establish participants’ journaling habits. Consequently, effects attributable solely to the baseline cannot be isolated from subsequent treatment effects. We also modeled introduction timing only as a binary factor (early vs. late). Future research could examine three or more introduction points to locate the balance between cognitive assistance and organic skill development, and to reveal growth trajectories not evident here. This study prioritized establishing overall effectiveness and did not model how individual differences shaped outcomes. For example, we did not examine whether work style, expertise, or intolerance of uncertainty moderated the tool’s utility. Analyzing these characteristics as moderators or mediators could inform more precise, personalized usage strategies. Such analyses would clarify for whom the tool is most effective and how prescriptions should vary by trait profile.

In terms of accessibility, the system was designed and evaluated for professional actors familiar with typical digital interfaces; therefore, accessibility features such as assistive-technology compatibility or support for varied digital literacy levels were not included. These considerations will be addressed in future versions as the system is prepared for wider deployment. Lastly, as the study was conducted entirely in Korean, findings should be interpreted within this linguistic and cultural context.

Post-study interviews also surfaced design improvements. For instance, the previously discussed “shared diary” feature merits exploration. It will be important to study how AI can support handoffs from solo work to ensemble collaboration. Expanding user-adjustable parameters (e.g., question count, theme) could better fit diverse creative preferences. Because some participants wanted more interactive dialogue beyond \textit{Actor’s Note} simply presenting questions, future versions could provide conversational feedback; subsequent studies could experimentally vary interaction richness to evaluate efficacy.

%% file: section/8Conclusion.tex
Positioning AI as a maieutic partner rather than a text-producing collaborator, this work shows that lightweight, context- and stage-aware scaffolding can make character journaling more feasible and more rewarding. In our in-the-wild evaluation, actors reported lower effort; stronger motivation and confidence; and more self-referential, emotionally expressive journals. Crucially, the timing of AI support shapes its function: early deployment jump-starts momentum, whereas later deployment deepens narrative engagement. These insights translate into concrete guidance for integrating AI into actor training: use selective prompts to lower initial friction without seizing authorship; tune intensity and focus as rehearsal progresses; and pair AI’s steady, nonjudgmental questioning with human partners’ improvisational spark. Framed this way, AI becomes a coach and sounding board—supporting, not supplanting, interpretive labor. Our findings sit alongside acknowledged constraints and point to next steps: adaptive, trait-aware personalization; conversational feedback; and collaborative modes (e.g., shared diaries) that bridge individual practice and ensemble work. Taken together, the paper advances a design stance for creativity support systems that preserves agency while sustaining reflective practice in performance training.

%% file: section/appendix.tex
\appendix
\section{Exemplar Prompts Used in Actor's Note}
\label{appendix:A} 

\subsection{Character Extraction Prompt}

\noindent\textbf{[System Role]} You are an expert Dramaturgical Analyst assisting actors in script preparation. Your objective is to parse the uploaded script and extract a structured profile of the primary characters.

\noindent\textbf{[Input Data]}
\begin{itemize}[noitemsep]
    \item Context: Full script text (uploaded file)
\end{itemize}

\noindent\textbf{[Rules and Constraints]}
\begin{enumerate}
    \item \textbf{Target Selection:} Identify up to 10 major characters who actively drive the narrative.
    \item \textbf{Script-Grounded Accuracy:} Extract information strictly from the provided text. Do not hallucinate or infer details not explicitly supported by dialogue or stage directions.
    \item \textbf{Name Fidelity:} Preserve character names exactly as written in the script (including original foreign spellings).
    \item \textbf{Concise Description:} For each character, summarize role, occupation (if stated), personality traits, and key narrative functions in 1--2 sentences.
    \item \textbf{No Meta Output:} Return only the structured list. Exclude any conversational filler, introductions, or closing remarks.
\end{enumerate}

\noindent\textbf{[Output Schema]}
\begin{itemize}[noitemsep]
    \item \textbf{Name:} [Character Name]
    \begin{itemize}
        \item \textbf{Profile:} [Role description, core traits, and narrative]
    \end{itemize}
    \item \textbf{Name:} [Character Name]
    \begin{itemize}
        \item \textbf{Profile:} [Role description, core traits, and narrative function]
    \end{itemize}
\end{itemize}

\subsection{Journal Topic Recommendation Prompt}

\noindent\textbf{[System Role]} You are a Maieutic Partner for actor training. Your goal is to facilitate deep character immersion through the Socratic method. You must support the actor's self-reflective inquiry without dictating interpretations or generating creative content on their behalf.

\noindent\textbf{[User Inputs]}
\begin{itemize}[noitemsep]
    \item Role Name: [Input Role Name]
    \item Role Information: [Input Role Description]
    \item Current Rehearsal Stage: [Input Rehearsal Stage]
\end{itemize}

\noindent\textbf{[Guiding Principles]}
\begin{itemize}
    \item \textbf{Question-Only Output:} Generate questions only. Do not provide explanations, interpretations, or advice.
    \item \textbf{Specificity over Abstraction:} Avoid vague or philosophical prompts. Each question should be concrete and directly answerable.
    \item \textbf{Actor Agency:} Do not imply correct answers or preferred interpretations.
    \item \textbf{Blank-Page Reduction:} Questions should lower the cognitive burden and help the actor begin writing immediately.
\end{itemize}

\noindent\textbf{[Stage-Aware Questioning Logic]} Adapt questions based on the Current Rehearsal Stage provided in the input:
\begin{itemize}
    \item \textbf{IF Script Analysis / Table Work:} Focus on facts, backstory, motivations, relationships, and given circumstances.
    \item \textbf{IF Standing Reading / Pre-blocking:} Focus on psychological reactions, justification for actions, and relational dynamics.
    \item \textbf{IF Scene Detail Work:} Focus on speech patterns, habits, attitudes, subtle emotional shifts, and external behaviors.
    \item \textbf{IF Run-through:} Focus on emotional continuity, plausibility of backstory, and internal consistency.
    \item \textbf{IF Performance / Other:} Integrated reflection across emotion, history, relationships, and action.
\end{itemize}

\noindent\textbf{[Question Categories]} Select and combine from the following types:
\begin{itemize}
    \item \textbf{Character Concretization} (e.g., “What small habit do you repeat when you feel anxious?”)
    \item \textbf{Emotional Exploration} (e.g., “What emotion arrives first before you speak this line?”)
    \item \textbf{Backstory Completion} (e.g., “What memory from before the play still influences this choice?”)
    \item \textbf{Relationships \& Change} (e.g., “How has your view of this character shifted since the last scene?”)
    \item \textbf{Counterfactual or Extreme Scenarios} (e.g., “How would you respond if this trust were suddenly broken?”)
\end{itemize}

\noindent\textbf{[Output Constraints]}
\begin{itemize}[noitemsep]
    \item Generate exactly three questions.
    \item One question per line.
    \item No bullet explanations, headers, or commentary.
\end{itemize}

\noindent\textbf{[Output Format]}
\begin{itemize}[noitemsep]
    \item {} [Question 1]
    \item {} [Question 2]
    \item {} [Question 3]
\end{itemize}
\section{Participant Demographics}
\label{appendix:B}

\begin{table}[H]
\caption{Full Participant-level Demographic Information}
\label{tab:participant_demographics}
\centering
\begin{tabular}{l c l p{3.2cm} c c}
\toprule
\textbf{ID} & \textbf{Age} & \textbf{Gender} & \textbf{Domain} & \textbf{Exp.} & \textbf{Journal?} \\ 
\midrule
P1 & 29 & Male & Musical, Theater, Film & 5 & Yes \\
P2 & 22 & Female & Theater & 3 & No \\
P3 & 46 & Female & Theater, Musical & 15 & No \\
P4 & 20 & Female & Theater, Musical & 3 & Yes \\
P5 & 37 & Male & Musical & 4 & No \\
P6 & 32 & Male & Theater, TV show & 10 & Yes \\
P7 & 28 & Male & Theater, Film & 9 & Yes \\
P8 & 24 & Female & Film, Drama & 3 & Yes \\
P9 & 33 & Male & Theater & 1 & No \\
P10 & 22 & Female & Theater, Musical & 4 & Yes \\
P11 & 33 & Female & Film, TV show & 9 & No \\
P12 & 21 & Female & Theater, Musical & 6 & Yes \\
P13 & 22 & Female & Theater, Musical & 2 & Yes \\
P14 & 34 & Male & Musical, Theater, Film & 5 & Yes \\
P15 & 31 & Female & Theater & 10 & Yes \\
P16 & 36 & Female & Musical & 4 & Yes \\
P17 & 29 & Female & TV show & 5 & Yes \\
P18 & 32 & Male & Musical & 4 & Yes \\
P19 & 25 & Male & Theater & 10 & Yes \\
P20 & 32 & Male & Theater, Film & 7 & Yes \\
P21 & 32 & Female & Theater & 9 & Yes \\
P22 & 26 & Female & Theater, Film & 7 & Yes \\
P23 & 34 & Male & Film, TV, Commer. & 8 & No \\
P24 & 32 & Female & Film, TV show & 2 & Yes \\
P25 & 49 & Female & Film, TV show & 2 & No \\
P26 & 23 & Female & Musical & 15 & Yes \\
P27 & 27 & Female & Musical & 2 & No \\
P28 & 27 & Female & Film, TV, Theater & 5 & Yes \\
P29 & 36 & Female & Theater & 13 & No \\
\bottomrule
\multicolumn{6}{p{0.9\columnwidth}}{\footnotesize \textit{Note.} Exp.: Acting Experience (Yrs); Journal?: Prior journaling experience.}
\end{tabular}
\end{table}

\aptLtoX{\section{Statistical Analysis Results}
\label{appendix:C}

\begin{table*}[t]
\caption{Randomization Check for Each Factor}
\label{tab:randomization_check}
\centering
\begin{tabular*}{\textwidth}{@{\extracolsep{\fill}}l c c c c}
\toprule
\textbf{Factor} & \textbf{G1 M (SD)} & \textbf{G2 M (SD)} & \textbf{\textit{t}(27)} & \textbf{\textit{p}} \\
\midrule
Acting Confidence (AC) & 4.76 (1.33) & 4.43 (1.28) & 0.72 & .481 \\
Cognitive Burden (CB) & 3.56 (1.18) & 3.07 (1.15) & 1.22 & .233 \\
Intrinsic Motivation (IM) & 5.50 (1.09) & 5.50 (1.09) & 0.00 & 1.000 \\
Character Understanding (CU) & 5.27 (1.24) & 5.11 (1.19) & 0.35 & .731 \\
Character Identification (CID) & 4.93 (1.17) & 4.57 (1.12) & 0.90 & .377 \\
Narrative Transportation (NT) & 4.63 (1.06) & 4.11 (1.01) & 1.50 & .146 \\
\bottomrule
\end{tabular*}
\end{table*}}{\begin{table*}[t]
\section{Statistical Analysis Results}
\label{appendix:C}

\caption{Randomization Check for Each Factor}
\label{tab:randomization_check}
\centering
\begin{tabular*}{\textwidth}{@{\extracolsep{\fill}}l c c c c}
\toprule
\textbf{Factor} & \textbf{G1 M (SD)} & \textbf{G2 M (SD)} & \textbf{\textit{t}(27)} & \textbf{\textit{p}} \\
\midrule
Acting Confidence (AC) & 4.76 (1.33) & 4.43 (1.28) & 0.72 & .481 \\
Cognitive Burden (CB) & 3.56 (1.18) & 3.07 (1.15) & 1.22 & .233 \\
Intrinsic Motivation (IM) & 5.50 (1.09) & 5.50 (1.09) & 0.00 & 1.000 \\
Character Understanding (CU) & 5.27 (1.24) & 5.11 (1.19) & 0.35 & .731 \\
Character Identification (CID) & 4.93 (1.17) & 4.57 (1.12) & 0.90 & .377 \\
Narrative Transportation (NT) & 4.63 (1.06) & 4.11 (1.01) & 1.50 & .146 \\
\bottomrule
\end{tabular*}
\end{table*}
}

\begin{table*}[h]
  \begin{minipage}[t]{0.49\textwidth}
    \caption{Normality Check for Study Variables}
    \label{tab:normality_check}
    \centering
    \begin{tabular*}{\linewidth}{@{\extracolsep{\fill}}l c c c}
    \toprule
    \textbf{Factor} & \textbf{S-W} & \textbf{\textit{p}} & \textbf{Result} \\
    \midrule
    Acting Confidence & 0.978 & .124 & Passed \\
    Cognitive Burden & 0.981 & .215 & Passed \\
    Intrinsic Motivation & 0.969 & .064 & Passed \\
    \bottomrule
    \end{tabular*}
  \end{minipage}
  \hfill 
  \begin{minipage}[t]{0.49\textwidth}
    \caption{Linearity and Homoscedasticity Check}
    \label{tab:linearity_check}
    \centering
    \begin{tabular*}{\linewidth}{@{\extracolsep{\fill}}l c c}
    \toprule
    \textbf{Relationship} & \textbf{Statistic} & \textbf{Result} \\
    \midrule
    Baseline--Treatment Linearity & $F(1,27)=42.3$ & Passed \\[2.8ex] 
    Residual Homoscedasticity & $BP=2.15$ & Passed \\
    \bottomrule
    \end{tabular*}
  \end{minipage}
\end{table*}

\begin{table*}[t]
\caption{Carryover Effects Analysis (Welch's T-test)}
\label{tab:carryover_analysis}
\centering
\begin{tabular*}{\textwidth}{@{\extracolsep{\fill}}l c l c c}
\toprule
\textbf{Factor} & \textbf{$\Delta$} & \textbf{\textit{t}(df)} & \textbf{\textit{p}} & \textbf{Carryover} \\
\midrule
Acting Confidence (AC) & 1.66 & $t(49.3)=3.88$ & $<.001^{***}$ & Yes \\
Cognitive Burden (CB) & -1.22 & $t(48.1)=-2.31$ & $.025^{*}$ & Yes \\
Intrinsic Motivation (IM) & 0.82 & $t(48.4)=2.09$ & $.042^{*}$ & Yes \\
Character Understanding (CU) & 0.029 & $t(45.6)=0.07$ & $.946$ & No \\
Character Identification (CID) & -0.042 & $t(50.1)=-0.08$ & $.934$ & No \\
Narrative Transportation (NT) & 0.033 & $t(52.0)=0.08$ & $.939$ & No \\
\bottomrule
\multicolumn{5}{l}{\footnotesize \textit{Note.} $\dagger p<.10, * p<.05, ** p<.01, *** p<.001$.}
\end{tabular*}
\end{table*}

\begin{table*}[t] 
\caption{Two-way ANOVA Results: Main and Interaction Effects of Group and Time across Outcome Variables.}
  \label{tab:anova_results}
  \centering
  \begin{tabular*}{\textwidth}{@{\extracolsep{\fill}}l c c c}
    \toprule
    \textbf{Variable} & \shortstack{\textbf{Group} \\ \small{$(df=1,27)$}} & \shortstack{\textbf{Time} \\ \small{$(df=2,54)$}} & \shortstack{\textbf{Group $\times$ Time} \\ \small{$(df=2,54)$}} \\
    \midrule
    Acting Confidence & $F=5.68$ $(p=.024^*)$ & $F=6.58$ $(p=.003^{**})$ & $F=2.15$ $(p=.127)$ \\
    \textbf{Cognitive Burden} & $F=0.31$ $(p=.580)$ & $F=2.28$ $(p=.112)$ & $\mathbf{F=7.15}$ $(\mathbf{p=.002^{**}})$ \\
    Character Understanding & $F=0.10$ $(p=.760)$ & $F=2.74$ $(p=.074^\dagger)$ & $F=0.72$ $(p=.492)$ \\
    Character Identification & $F=0.49$ $(p=.491)$ & $F=5.24$ $(p=.008^{**})$ & $F=0.68$ $(p=.511)$ \\
    Intrinsic Motivation & $F=1.14$ $(p=.294)$ & $F=1.66$ $(p=.201)$ & $F=1.73$ $(p=.186)$ \\
    \textbf{Narrative Transportation} & $F=1.61$ $(p=.216)$ & $F=3.31$ $(p=.044^*)$ & $\mathbf{F=2.45}$ $(\mathbf{p=.096^\dagger})$ \\
    \bottomrule
    \multicolumn{4}{l}{\footnotesize \textit{Note.} $\dagger p < .10, * p < .05, ** p < .01, *** p < .001$. Bold indicates significant or marginal interactions.}
  \end{tabular*}
\end{table*}

\begin{table*}[t]
\caption{Pairwise Comparisons of Cognitive Burden (CB), Acting Confidence (AC), and Narrative Transportation (NT)}
\label{tab:pairwise_comparisons}
\centering
\begin{tabular*}{\textwidth}{@{\extracolsep{\fill}}l l l l}
\toprule
\textbf{Factor} & \textbf{Comparison} & \textbf{\textit{p}-value} & \textbf{Cohen's \textit{d}} \\
\midrule
\textbf{Cognitive Burden (CB)} & Early - Late at T1 & .2106 & $+0.47 (d_s)$ \\
 & Early - Late at T2 & $.0060^{**}$ & $-1.06 (d_s)$ \\
 & Early - Late at T3 & .7600 & $+0.11 (d_s)$ \\
 & Early-AI T1 - T2 & .5161 & $+0.20 (d_z)$ \\
 & Early-AI T1 - T3 & .8984 & $-0.08 (d_z)$ \\
 & Early-AI T2 - T3 & .2784 & $-0.29 (d_z)$ \\
 & \textbf{Late-AI} T1 - T2 & $.0006^{***}$ & $-1.02 (d_z)$ \\
 & \textbf{Late-AI} T1 - T3 & .2548 & $-0.41 (d_z)$ \\
 & \textbf{Late-AI} T2 - T3 & $.0561^{\dagger}$ & $+0.61 (d_z)$ \\
\midrule
\textbf{Acting Confidence (AC)} & Early - Late at T1 & .4016 & $+0.31 (d_s)$ \\
 & Early - Late at T2 & $.0049^{**}$ & $+1.09 (d_s)$ \\
 & Early - Late at T3 & $.0475^{*}$ & $+0.75 (d_s)$ \\
 & \textbf{Early-AI} T1 - T2 & $.0371^{*}$ & $-0.68 (d_z)$ \\
 & \textbf{Early-AI} T1 - T3 & $.0032^{**}$ & $-0.92 (d_z)$ \\
 & \textbf{Early-AI} T2 - T3 & .6429 & $-0.24 (d_z)$ \\
 & \textbf{Late-AI} T1 - T2 & .9061 & $+0.11 (d_z)$ \\
 & \textbf{Late-AI} T1 - T3 & .2171 & $-0.44 (d_z)$ \\
 & \textbf{Late-AI} T2 - T3 & $.0960^{\dagger}$ & $-0.55 (d_z)$ \\
\midrule
\textbf{Narrative Trans. (NT)} & Early - Late at T1 & .1496 & $+0.54 (d_s)$ \\
 & Early - Late at T2 & $.0774^{\dagger}$ & $+0.67 (d_s)$ \\
 & Early - Late at T3 & .9632 & $+0.02 (d_s)$ \\
 & \textbf{Early-AI} T1 - T2 & .3760 & $-0.36 (d_z)$ \\
 & \textbf{Early-AI} T1 - T3 & .8023 & $-0.17 (d_z)$ \\
 & \textbf{Early-AI} T2 - T3 & .7570 & $+0.19 (d_z)$ \\
 & \textbf{Late-AI} T1 - T2 & .7421 & $-0.19 (d_z)$ \\
 & \textbf{Late-AI} T1 - T3 & $.0128^{*}$ & $-0.76 (d_z)$ \\
 & \textbf{Late-AI} T2 - T3 & $.0779^{\dagger}$ & $-0.57 (d_z)$ \\
\bottomrule
\multicolumn{4}{l}{
  \footnotesize \textit{Note.} $^\dagger .05 \le p < .10, ^* p < .05, ^{**} p < .01, ^{***} p < .001$. Effect sizes are Cohen's $d$.
} \\
\multicolumn{4}{l}{
  \footnotesize $d_s$: independent-samples $d$ (Early vs. Late, pooled SD); $d_z$: paired-samples $d$ (within-group comparisons).
}
\end{tabular*}
\end{table*}